\documentclass{article}
\usepackage[T1]{fontenc}
\usepackage[utf8]{inputenc}

\usepackage[colorlinks=true, citecolor=blue, linkcolor=blue, urlcolor=blue]{hyperref}
\usepackage{amsmath}
\usepackage{amssymb}
\usepackage{graphicx}
\usepackage{booktabs}

\usepackage[
  backend=biber,
  style=authoryear,
  dashed=false,
  maxcitenames=2,
  uniquename=false,
  maxbibnames=99,
  giveninits=true
]{biblatex}

\DeclareDelimFormat{nameyeardelim}{\addcomma\space}

\usepackage[disable]{todonotes}

\usepackage{geometry}
\usepackage{titlesec}
\titleformat{\section}{\normalfont\large\bfseries}{\thesection}{1em}{}
\titleformat{\subsection}{\normalfont\normalsize\bfseries}{\thesubsection}{1em}{}
\usepackage[font=footnotesize, labelfont=bf, margin=1cm]{caption}

\usepackage{authblk}

\graphicspath{ {./images/} }
\usepackage{subcaption}

\usepackage{amsthm}

\title{\textbf{Bayes Factor Testing of Linear and Nonlinear Effects in Additive Gaussian Process Regression}}
\author{Yueheng Wu}
\author{Joran Jongerling}
\author{Joris Mulder}
\affil{Department of Methodology and Statistics, Tilburg University}

\date{}

\begin{document}
\maketitle

\begin{abstract}
Gaussian process regression flexibly models nonlinear relationships between predictors and an outcome variable. A central question in this modeling framework is whether each predictor affects the outcome linearly, nonlinearly, or not at all. We propose a Bayesian testing framework for this problem within additive Gaussian process regression. Each predictor is decomposed into a standardized linear component with a $g$-prior and an orthogonalized Gaussian process component capturing nonlinear structure. The latter is governed by a standardized magnitude parameter whose value zero removes the component. Local and non-local priors are considered for the standardized magnitude parameter. Bayes factors for testing the linear and nonlinear effects are computed from the fitted full model via Savage--Dickey-type identities, avoiding expensive marginal likelihood computations of many different models. To assess the accuracy of the Bayes factor estimates, the outcomes are compared to bridge sampling estimates. Simulations and empirical applications show the performance and applicability of the methodology.
\end{abstract}

\section{Introduction}

Many real-world relationships are not well described by a straight line. Identifying and characterizing such nonlinear effects is therefore a recurring problem in applied research. When the relationship is close to linear, linear regression remains useful due to its parsimonious and computationally simple solution and the simple interpretation of the effect sizes. When the true nature of the relationships is unknown, however, a key question is whether the relationships are zero, linear, or nonlinear. The goal of the present paper is to address this question separately for each predictor.

Within the linear modeling framework, nonlinearity can be added through pre-specified nonlinear basis expansions, such as polynomial terms or regression splines \parencite{stone1997use,levelt2026modelling}.
However, these approaches require the researcher to choose a functional form in advance, and are therefore not recommended when the true nonlinear shapes are unknown.

Gaussian process (GP) regression \parencite{williams2006gaussian} provides a kernel-based framework for modeling unknown relationships between predictors and a response variable. Rather than fixing a specific parametric form, a Gaussian process places a prior directly on the function relating predictors to the response, with smoothness assumptions encoded through a structured covariance kernel typically having a magnitude parameter, quantifying the size of the nonlinear effect, and a length-scale parameter, quantifying the smoothness of the nonlinear shape \parencite{Williams1996}. The data then inform the shape of the nonlinear function. This makes GP regression a flexible alternative when the forms of the predictor--response relationships are not known in advance.

Despite its flexibility, determining whether a predictor has an effect on the outcome is less straightforward in GP regression than in linear regression. The earliest approach to this problem, automatic relevance determination (ARD), uses predictor-specific length-scale parameters as a proxy for its relevance \parencite{mackay1992bayesian, neal1995bayesian}. As the length-scale parameter quantifies the smoothness of the function, a very large length-scale, implying a highly smooth function, could reflect a low influence on the response. On the other hand, it has also been shown that a small length-scale may actually imply a small association between the predictor and the outcome \parencite{linkletter2006variable, nakajima2007theoretical}, and that ARDs may not be useful when an effect may be large but close to linear \parencite{piironen2016projection}.

Sensitivity analysis and projection-based variable selection assess predictor influence by measuring predictive change. Sensitivity-based approaches summarize how much the posterior predictive distribution changes when an input is perturbed, either globally over the input domain by attributing model-output variation to individual predictors or their interactions \parencite{oakley2004probabilistic}, or locally near the observed data \parencite{paananen2019variable}. Projection-based methods instead fit a reference model on all predictors and then identify reduced models whose predictive distributions remain close to that reference \parencite{piironen2016projection}. Both families describe predictive influence, but their target is predictive change or fidelity rather than evidence for a null hypothesis.

Bayesian sparsity priors address variable selection more directly by encoding inclusion decisions within the prior \parencite{mitchell1988bayesian, george1993variable}. In GP models, this can be achieved with spike-and-slab formulations or binary inclusion indicators on inputs, kernel parameters, or covariance components \parencite{linkletter2006variable, savitsky2011, titsias2011spike}, which assign posterior probability to whether a predictor or component is active. These approaches are closer to inferential variable selection than sensitivity or projection methods, but posterior inclusion probabilities require prior inclusion probabilities which may be difficult to elicit. 

A fundamental difficulty is that linear and nonlinear contributions are not separated within the previously mentioned approaches, nor in standard GP inference. \textcite{gramacyGaussianProcessesLimiting2008} showed that the linear model is a limiting special case of the GP, yet standard GP priors assign negligible posterior probability to this limit even when the data are genuinely linear, so a fitted GP conflates the two types of structure. This motivates an approach that makes the separation explicit. 

One option is kernel design strategies, which decompose the covariance function into structured components corresponding to main effects and interactions, as in additive and functional-ANOVA kernel constructions \parencite{duvenaud2011a, durrande2013anova}. Structure-discovery approaches extend this idea by searching over sums and products of kernels to obtain expressive but interpretable covariance structures \parencite{duvenaud2013structure}, and hierarchical priors or kernel-weighting schemes can encourage sparsity across components \parencite{archambeau2011multiple}. These approaches provide flexible mechanisms for estimating linear and nonlinear structure, but estimation does not constitute hypothesis testing or model comparison. In particular, when the presence of a kernel component is governed by a nonnegative amplitude or variance parameter, its absence corresponds to a boundary value of the parameter space, making standard Wald- or likelihood-ratio-based inference problematic. 

As an alternative, Bayes factors allow statistical inference via the marginal likelihoods of competing models, quantifying the relative evidence they contain while penalizing unnecessary complexity \parencite{kass1995bayes, jeffreys1961theory}. Marginal likelihoods for GP regression involve analytically intractable integrals, and although Bayes factors for GP models have been studied \parencite{mukhopadhyay2021}, their practical use has been limited to simple testing problems \parencite{mulder2023}.

The present paper targets the presence of linear and nonlinear contributions through a predictor-specific magnitude parameter in a multi-predictor additive GP model. For each predictor, the model contains linear components using mixtures of $g$-priors \parencite{ZellnerSiow1980, Liang2008} and a nonlinear Gaussian process component constructed to be orthogonal to the linear space \parencite{PlumleeJoseph2018}, so that the nonlinear component captures only structure not already represented by the linear effect. The magnitude of each nonlinear component is controlled by a standardized scale parameter (relative to the scale of the noise), similar to the modeling of standardized linear effects \parencite{Rouder2012}. As the corresponding nonlinear contribution vanishes regardless of the smoothness of the underlying Gaussian process, testing whether a nonlinear effect is present reduces to testing whether the standardized magnitude parameter is zero.

The standardized parameterization yields a clear scale of the degree of nonlinearity, which simplifies the specification of the prior and its scale. Different priors are considered including a half $t$ prior and non-local priors \parencite{johnson2010nonlocal}. For non-local priors, the prior density vanishes at the null value and thereby sharpens the separation between the null where nonlinear effect is absent and the alternative where the nonlinear effect is present. Evidence for absence/presence of each linear and nonlinear effect is quantified through Bayes factors obtained via Savage--Dickey-type identities \parencite{Dickey1971, verdinelli1995, wetzels2010encompassing, mulder2022generalization,WilliamsMulder2020JMP}. This improves computational feasibility as a single fit of the full model is only required for testing all separate nonlinear and linear elements. This is particularly useful due to the computational burden when fitting GPs. In contrast, approaches that evaluate each hypothesis through its own marginal likelihood, such as bridge sampling \parencite{Meng1996, Gronau2017} or sensitivity-based model comparison, require a separate model fit for each subset model, so that the number of required fits grows with the number of predictors. Bridge sampling is used as an independent benchmark. The behavior of the procedure is examined through a simulation study, and its practical use is illustrated on real datasets with comparison to findings from previous analyses.

The remainder of the paper is organized as follows. Section~\ref{sec:model} develops the additive decomposition of each predictor into orthogonal linear and nonlinear components, and Section~\ref{sec:testing} formulates the associated hypothesis tests. Section~\ref{sec:priors} specifies the priors on the magnitude parameters, and Section~\ref{subsec:bf-computation} describes how the linear and nonlinear Bayes factors are computed from a single model fit. Section~\ref{sec:simulation} reports a simulation study, and Section~\ref{sec:empirical} illustrates the method on three datasets. Section~\ref{sec:discussion} concludes.

\section{Modeling framework\label{sec:model}}

Let $\mathbf{y}\in\mathbb{R}^n$ denote a vector of $n$
observations of the response variable $Y$, and let
$\mathbf{X}\in\mathbb{R}^{n\times p}$ be the full-rank design matrix
of the $p$ candidate predictors, each centered at zero and rescaled to have range one.
We model the relationship between predictors and the response through
an additive decomposition of linear and nonlinear contributions:
\begin{align}
\label{eq:model}
\mathbf{y}
  &=
  \alpha \mathbf{1}_n
  + \sigma\,\mathbf{X}\boldsymbol{\theta}
  + \sum_{j=1}^{p}\mathbf{f}^{\perp}_j(\mathbf{x}_j)
  + \boldsymbol{\varepsilon}, \\ 
  \label{eq:error}
\boldsymbol{\varepsilon}
  &\sim \mathcal{N}(\mathbf{0},\,\sigma^2\mathbf{I}_n)
\end{align}
where $\boldsymbol{\theta}=(\theta_1,\dots,\theta_p)^\top$ collects
the standardized linear coefficients, via the common scale factor $\sigma$, each
$\mathbf{f}^{\perp}_j(\mathbf{x}_j)\in\mathbb{R}^n$ is a latent nonlinear
component associated with predictor $j$, constructed from orthogonal Gaussian processes (discussed below), and
$\boldsymbol{\varepsilon}$ is a Gaussian error term with unknown
variance $\sigma^2$. The common scale factor $\sigma$ ensures that the linear 
effects lie on a standardized scale relative to the residual scale. For the nonlinear component, a standardized parameterization will also be proposed.

\subsection{Nonlinear component: Gaussian process prior
\label{subsec:gp}}

To represent the nonlinear effects, we start with a Gaussian process prior
on latent function $f_j(\cdot)$, for $j=1,\ldots,p$.
A Gaussian process defines a distribution of random functions
$\{f_j(x):x\in\mathcal{X}\}$ such that every finite collection of
function evaluations has a multivariate normal distribution.
Thus, the GP defines a stochastic process over the predictor domain,
while its evaluations at the $n$ observed predictor values form an
$n$-dimensional multivariate normal distribution.

We write
\[
\mathbf{f}_j(\mathbf{x}_j)
=
\bigl(f_j(x_{1j}),\dots,f_j(x_{nj})\bigr)^\top
\]
to denote the GP realization at the observed values of predictor $j$.
A GP is fully characterized by a mean function and a covariance kernel:
the mean function describes the prior average trend, whereas the kernel
determines how strongly function values at two input locations are
expected to co-vary.
In the present model, we set the GP mean function to zero because the
intercept and linear effects are already represented explicitly by the
regression components,
$\alpha\mathbf{1}_n+\sigma\,\mathbf{X}\boldsymbol{\theta}$.

For each predictor $j$, we use the squared exponential kernel
\begin{equation}
\label{eq:se-kernel}
k_j(x,x'\mid\lambda_j)
  =
  \exp\!\left(
      -\frac{(x-x')^2}{2\lambda_j^2}
    \right),
\end{equation}
which satisfies $k_j(x,x)=1$. This kernel implies smooth sample paths: larger length-scales ($\lambda_j$) produce
more slowly varying nonlinear trends, whereas smaller values of $\lambda_j$
allow more local wiggliness. Other kernels, such as those from the Mat\'ern class, could be used
when rougher sample paths are substantively more appropriate (we come back to this in the Discussion).
The corresponding kernel matrix $\mathbf{K}_j$ is obtained by
evaluating $k_j(\cdot,\cdot\mid\lambda_j)$ at all observed pairs of
predictor values.

Next we model the nonlinear components using a standardized magnitude parameter $\delta_j$. This is achieved by including the error standard deviation $\sigma$ in the kernel as follows
\begin{equation}
\label{eq:gp-finite}
\mathbf{f}_j(\mathbf{x}_j)
\mid \delta_j,\lambda_j
\sim
\mathcal{N}\!\bigl(
\mathbf{0},\;
\sigma^2\,\delta_j^{\,2}\mathbf{K}_j
\bigr).
\end{equation}

\begin{figure}[htbp] \centering \includegraphics[width=1\textwidth]{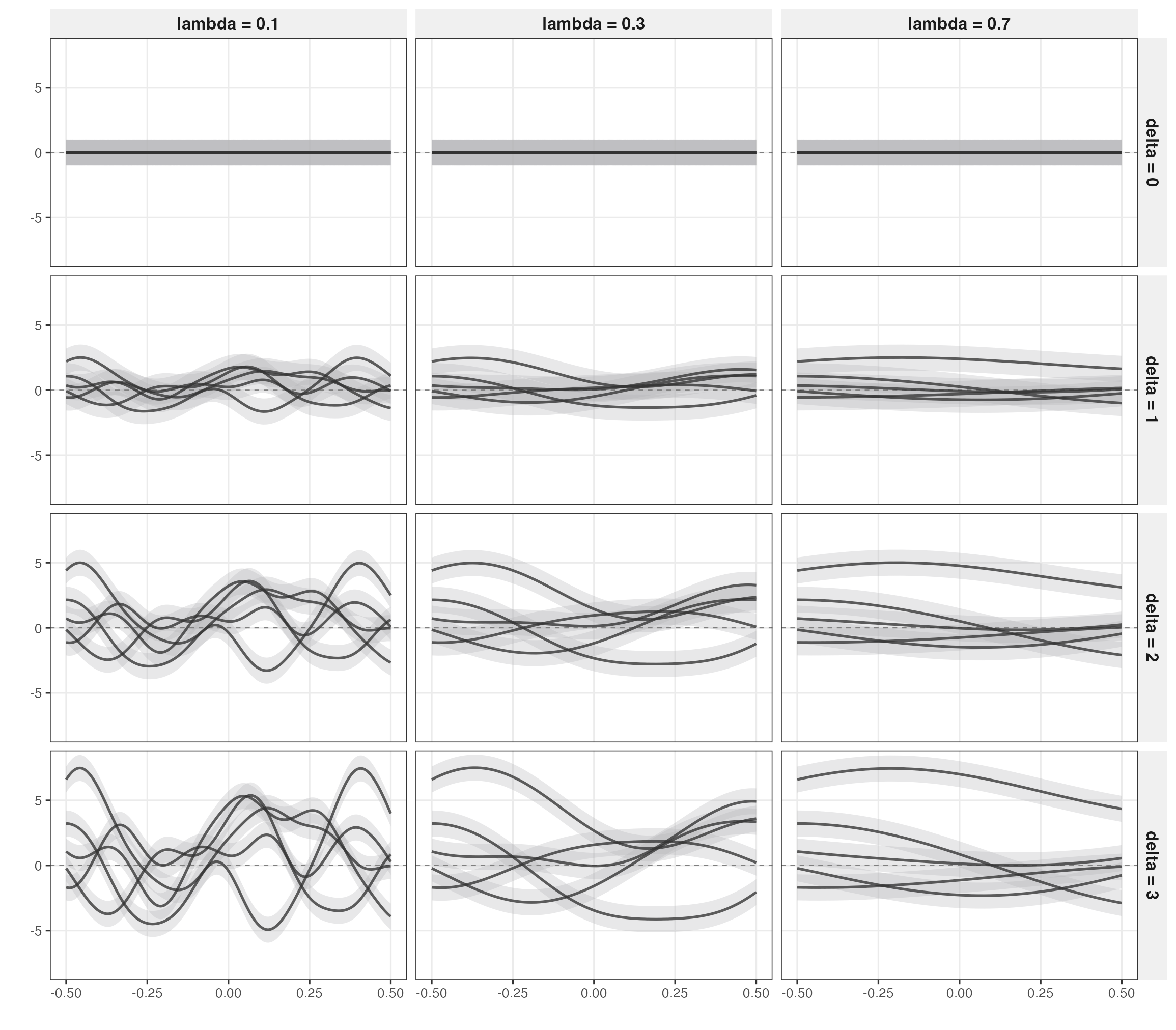} \caption{Gaussian process realizations for selected values of the standardized nonlinear magnitude $\delta$ and the length-scale $\lambda$, with the predictor range fixed at $1$ and the error standard deviation fixed at $\sigma=1$. Each panel shows five realizations generated from a common random seed. The grey ribbon around each realization spans $f(x)\pm\sigma$ and marks the scale of the observation noise.}
\label{fig:gp_effects}
\end{figure}

To better understand the roles of the two GP hyperparameters, we plotted several random functions for different combinations of hyperparameters in Figure~\ref{fig:gp_effects}.
Each realization is shown together with a grey ribbon spanning $f(x)\pm\sigma$, which marks the scale of the observation noise. Because the error standard deviation is held fixed at $\sigma=1$ across all panels, this ribbon has the same width throughout and serves as a common reference, so that the realizations can be compared directly across panels and $\delta_j$ retains the same meaning for each combination of hyperparameters. As $\delta_j$ decreases toward zero, the nonlinear contribution vanishes regardless of the value of $\lambda_j$. The magnitude parameter therefore captures the relative importance of the nonlinear structure.

\subsection{Orthogonalization of the nonlinear component
\label{subsec:orthogonalization}}

A key difficulty is that the GP terms can absorb variation that is
already explained by the linear predictor, because smooth GP
realizations can themselves exhibit broad trends over the observed
predictor range, and with large $\lambda_j$ these trends can be close
to linear.
This creates confounding between the linear and nonlinear parts of the
model, making the decomposition hard to interpret.

Following \textcite{PlumleeJoseph2018}, we therefore orthogonalize each
GP realization with respect to the space spanned by the parametric mean function. We write the augmented design
matrix $\tilde{\mathbf{X}}=[\mathbf{1}~\mathbf{X}]$, which combines the intercept and the linear predictor terms.
Let
\begin{equation}
\label{eq:projection}
\mathbf{P}
  =
  \tilde{\mathbf{X}}
  (\tilde{\mathbf{X}}^{\!\top}\tilde{\mathbf{X}})^{-1}
  \tilde{\mathbf{X}}^{\!\top}
\end{equation}
be the projection matrix onto $\operatorname{col}(\tilde{\mathbf{X}})$.
We define the orthogonalized nonlinear component as
\begin{equation}
\label{eq:orthogonalized-f}
\mathbf{f}_j^{\perp}(\mathbf{x}_j)
=
(\mathbf{I}_n-\mathbf{P})\mathbf{f}_j(\mathbf{x}_j).
\end{equation}
It follows that the corresponding covariance matrix is
\begin{equation}
\label{eq:ortho-kernel}
\mathbf{K}_j^{\perp}
  =
  (\mathbf{I}_n-\mathbf{P})
  \mathbf{K}_j
  (\mathbf{I}_n-\mathbf{P}),
\end{equation}
so that
\begin{equation}
\label{eq:prior-fj}
\mathbf{f}_j^{\perp}(\mathbf{x}_j)
\mid \delta_j,\lambda_j
\sim
\mathcal{N}\!\bigl(
\mathbf{0},\;
\sigma^2\,\delta_j^{\,2}\mathbf{K}_j^{\perp}
\bigr).
\end{equation}
This construction ensures that the nonlinear component is orthogonal
to the linear predictor space, so that the GP captures only structure
not already represented by the intercept and linear terms.

\subsection{Marginal likelihood formulation\label{subsec:marginal}}

Because the latent vectors
$\mathbf{f}^{\perp}_1(\mathbf{x}_1),\dots,\mathbf{f}^{\perp}_p(\mathbf{x}_p)$ are
Gaussian, they can be integrated out analytically.
This yields a marginal sampling distribution under the full model,
which contains both the linear component and all additive nonlinear
components in the covariance matrix:
\begin{equation}
\label{eq:marginal}
M_1:\quad
\mathbf{y}\mid\alpha,\sigma,\boldsymbol{\theta},
\boldsymbol{\delta},\boldsymbol{\lambda}
\sim
\mathcal{N}\!\Bigl(
  \alpha\mathbf{1}_n
  + \sigma\,\mathbf{X}\boldsymbol{\theta},\;
  \sigma^2\mathbf{I}_n
  + \sigma^2\sum_{j=1}^{p}
    \delta_j^{\,2}\mathbf{K}^{\perp}_j
\Bigr),
\end{equation}
where $\boldsymbol{\delta}=(\delta_1,\dots,\delta_p)$ and
$\boldsymbol{\lambda}=(\lambda_1,\dots,\lambda_p)$.

This marginalized form is convenient because it avoids explicit
sampling of the high-dimensional latent functions while remaining
equivalent, for inference on $\mathbf{y}$, to the hierarchical
representation of the model.

\section{Hypothesis testing framework\label{sec:testing}}

For each predictor, we consider two inferential questions: whether its
nonlinear coefficient should be included, and whether its linear
component should be included.

\subsection{Testing nonlinear effects\label{subsec:testing-nonlinear}}

To test the nonlinear effect of predictor $j$, we compare the full
model against a restricted model in which the $j$th nonlinear
component is removed:
\begin{equation}
\label{eq:model-minus-delta-j}
M_{0,j}^{(N)}:\quad
\mathbf{y}\mid\alpha,\sigma,\boldsymbol{\theta},
\boldsymbol{\delta}_{-j},\boldsymbol{\lambda}_{-j}
\sim
\mathcal{N}\!\Bigl(
  \alpha\mathbf{1}_n
  + \sigma\,\mathbf{X}\boldsymbol{\theta},\;
  \sigma^2\mathbf{I}_n
  + \sigma^2\sum_{l\neq j}
    \delta_l^{\,2}\mathbf{K}^{\perp}_l
\Bigr).
\end{equation}
where $\boldsymbol{\delta}_{-j}$ and $\boldsymbol{\lambda}_{-j}$ indicate that no nonlinear effects (i.e., magnitude and length-scale) are estimated for the $j$th predictor and the summation $\sigma^2\sum_{l\neq j} \delta_l^{\,2}\mathbf{K}^{\perp}_l$ therefore also does not take predictor $j$ into account. 

This corresponds to testing
\begin{equation}
\label{eq:test-nonlinear}
H_{0,j}^{(N)}:\;\delta_j=0
\qquad\text{versus}\qquad
H_{1,j}^{(N)}:\;\delta_j>0.
\end{equation}

\subsection{Testing linear effects\label{subsec:testing-linear}}

To test the linear effect of predictor $j$, we compare the full model
against a restricted model in which the $j$th linear coefficient is
removed:
\begin{equation}
\label{eq:model-minus-theta-j}
M_{0,j}^{(L)}:\quad
\mathbf{y}\mid\alpha,\sigma,\boldsymbol{\theta}_{-j},
\boldsymbol{\delta},\boldsymbol{\lambda}
\sim
\mathcal{N}\!\Bigl(
  \alpha\mathbf{1}_n
  + \sigma\,\mathbf{X}_{-j}\boldsymbol{\theta}_{-j},\;
  \sigma^2\mathbf{I}_n
  + \sigma^2\sum_{l=1}^{p}
    \delta_l^{\,2}\mathbf{K}^{\perp}_l
\Bigr),
\end{equation}
where $\boldsymbol{\theta}_{-j}$ denotes the vector of standardized linear effects excluding the $j$-th effect and $\mathbf{X}_{-j}$ denotes the design matrix where the $j$-th column is excluded. We can write the hypothesis testing problem of the $j$-th linear component as follows
\begin{equation}
\label{eq:test-linear}
H_{0,j}^{(L)}:\;\theta_j=0
\qquad\text{versus}\qquad
H_{1,j}^{(L)}:\;\theta_j\neq0.
\end{equation}

\section{Prior specification\label{sec:priors}}

This section specifies the priors on the model parameters and
hyperparameters.

\subsection{Prior on the linear coefficients\label{subsec:prior-theta}}

For the linear coefficients $\boldsymbol{\theta}$, we use Zellner's
$g$-prior \parencite{Zellner1986}, a standard construction in Bayesian
regression:
\begin{equation}
\label{eq:prior-theta}
\boldsymbol{\theta}\mid g
  \sim
  \mathcal{N}\!\bigl(
    \mathbf{0},\;
    g\,(\mathbf{X}^{\!\top}\mathbf{X})^{-1}
  \bigr).
\end{equation}
The $g$-prior scales the covariance of
$\boldsymbol{\theta}$ by $(\mathbf{X}^{\!\top}\mathbf{X})^{-1}$,
so that the prior adapts to the correlation structure and geometry of
the predictors \parencite{Liang2008, Rouder2012}.
The hyperparameter $g$ determines the overall amount of shrinkage:
small values pull the linear coefficients toward zero, whereas large
values yield weaker shrinkage. The intercept $\alpha$ is common to all models under comparison and is therefore assigned a flat prior, following standard practice for $g$-priors \parencite{Liang2008, Bayarri2012}.

We assign an inverse-gamma hyperprior to the shrinkage parameter:
\begin{equation}
\label{eq:prior-g}
g
\sim
\mathrm{IG}\!\bigl(a=2,\;b=\tfrac{n}{2}\bigr).
\end{equation}
The scale $b=n/2$ is the Zellner--Siow scale
\parencite{ZellnerSiow1980, Liang2008}, which ties the prior on $g$ to
the sample size so that the shrinkage induced on
$\boldsymbol{\theta}$ remains stable as $n$ grows.
This specification belongs to the family of mixture-of-$g$ priors
studied by \textcite{Liang2008}, who note that the classical
Zellner--Siow Cauchy prior
\parencite{ZellnerSiow1980},
$g\sim\mathrm{IG}(1/2,n/2)$, does not permit closed-form marginal
likelihoods, limiting its adoption in practice.
In our joint linear--GP model the problem is more severe: with $a=1/2$
the prior mean of $g$ does not exist, and the central $98\%$ of the prior
spans more than four orders of magnitude, so that the conditional scale
$\sqrt{g\,v_j}$ of $\theta_j$ varies by a factor of roughly $200$ under
the prior. Such a funnel-shaped geometry cannot be traversed efficiently
with a single step size and often produces divergent transitions in
Hamiltonian Monte Carlo sampling \parencite{Stan}.
Setting $a=2$ reduces this spread to a factor of about $7$ on the scale
of $\theta_j$, while still yielding a proper prior with a
finite-variance $t(4)$ marginal on $\boldsymbol{\theta}$.

\subsection{Priors on \texorpdfstring{$\delta_j$}{delta j}
\label{subsec:prior-delta}}
The standardized GP magnitude $\delta_j$ controls the presence and
strength of the $j$th nonlinear effect.
Because $\delta_j$ is the ratio of the GP marginal standard deviation
to the noise standard deviation before orthogonalization, it has a
direct signal-to-noise interpretation.
This interpretation applies for all predictors, which allows the prior
on $\delta_j$ to be calibrated on a common and
interpretable scale.

We begin with a half Student-$t$ prior for the standardized nonlinear
magnitude:
\begin{equation}
\label{eq:delta_prior_t}
\delta_j\mid s_\delta
\sim
t^{+}_{\nu}(s_\delta),
\end{equation}
where $t^{+}_{\nu}(s)$ denotes the half Student-$t$ distribution with
degrees of freedom $\nu$ and scale $s$.
This choice follows Jeffreys' principle:
small deviations from the null may be more plausible a priori than
large ones, while the tails should remain sufficiently heavy to
accommodate larger effects when supported by the data
\parencite{jeffreys1961theory}.
It is also consistent with \textcite{gelman2006}, who recommends half
Student-$t$ priors as weakly informative priors for scale parameters
in hierarchical models.

The standardized interpretation of $\delta_j$ facilitates the choice of the prior scale.
With predictors rescaled to $\mathrm{range}(x_j)=1$,
Figure~\ref{fig:gp_effects} suggests that
$\delta\approx1,\,2,\,3$ correspond roughly to small, medium, and large
nonlinear departures on this scale.
We therefore use a prior median of $\delta=2$ as a default calibration,
representing a medium nonlinear effect.

\begin{figure}[htbp]
    \centering
    \includegraphics[width=\textwidth]{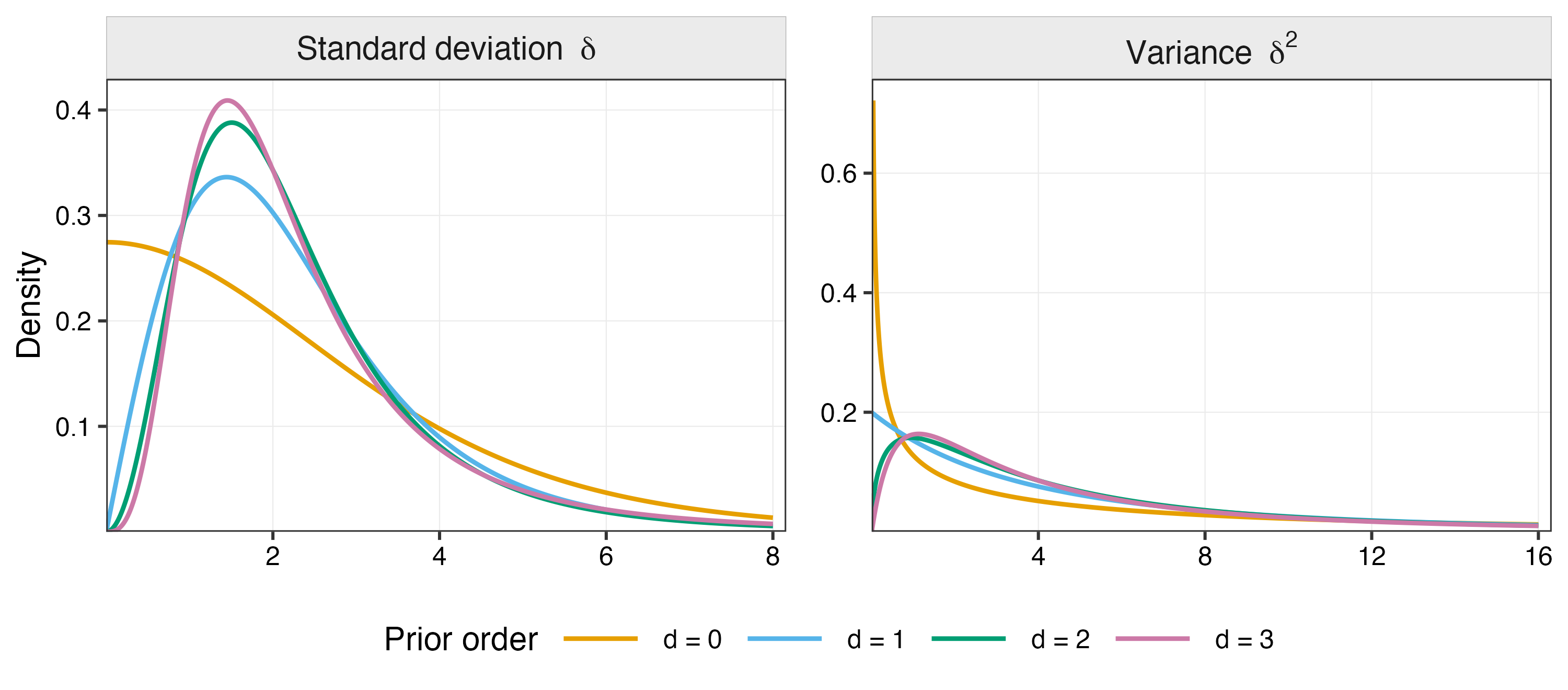}
    \caption{Prior densities for the standardized nonlinear magnitude
      $\delta$ (left panel) under the local prior ($d=0$, orange) and non-local
      priors of order $d=1$ (blue), $d=2$ (green), and $d=3$ (pink).
      All four priors are calibrated to a common median of
      $\delta=2$.
      As the order $d$ increases, the density vanishes more rapidly
      near $\delta=0$, yielding progressively sharper separation
      between the null hypothesis $H_0{:}\;\delta=0$ and the
      alternative $H_1{:}\;\delta>0$. The right panel displays the equivalent
      priors for the variances $\delta^2$, truncated at $\delta^2=16$; on this
      scale the local prior diverges as $\delta^2\to0$ and its density leaves
      the panel at the top.}
    \label{fig:delta-priors}
\end{figure}

In this paper, we take the half Student-$t$ prior
$t_6^{+}(2.79)$, where $\nu=6$ provides finite variance while retaining
moderately heavy tails and $s_\delta=2.79$ places the prior median at
$\delta=2$. Figure~\ref{fig:delta-priors} displays this prior density (orange lines). 
If smaller or larger nonlinear effects are expected a priori, the same
calibration can be shifted accordingly: under the half Student-$t$
prior, medians at $\delta=1$ and $\delta=3$ correspond to scale values
approximately $s_\delta=1.39$ and $s_\delta=4.18$, respectively.

Figure~\ref{fig:delta-priors} shows that the half Student-$t$ prior does not sharply separate the null ($\delta=0$) and the alternative ($\delta>0$) as considerable prior mass is placed on small values of
$\delta$. This can be clearly seen when viewing the half $t$ prior on the variance scale, showing an asymptote at $\delta=0$. This also illustrates why half t priors for prior standard deviations are popular for Bayesian regularization by constructing horseshoe-type priors \parencite{carvalho2010,polsonscott2012,MulderPericchi:2018}. In a testing problem on the other hand this local behavior around the null may not be preferred as the null and alternative may not be sufficiently distinguishable around the null, potentially understating the evidence for the null, which was the motivation to also consider non-local priors. 
Specifically, we consider the moment prior, introduced by
\textcite{johnson2010nonlocal}, for the standardized magnitude parameter. This can be done by multiplying a base
prior by a power of $\delta$ so that the prior density approaches zero when approaching
the null value.
In the present setting, this yields
\begin{equation}
\label{eq:nonlocal-prior}
\pi_d(\delta_j)
  =
  \frac{1}{C_d}\,
  \delta_j^{\,d}\,
  t_6^+\!\left(
    \delta_j;\frac{s_\delta}{\kappa}
  \right),
\qquad
\delta_j>0,
\end{equation}
where $C_d$ is a normalizing constant and $\kappa>0$ rescales the base
prior.
Thus, we take the half Student-$t$ prior as base
prior, so the moment prior is obtained by rescaling this prior through
$\kappa$ and then applying the tilt $\delta_j^d$.
The same construction could, in principle, be combined with other
base priors.
When $d=0$, the tilt reduces to unity, and the construction recovers
the local half Student-$t$ prior from the previous paragraph; in
particular, when $\kappa=1$, it coincides exactly with the default prior
$t_6^+(2.79)$.

\subsection{Prior on \texorpdfstring{$\lambda_j$}{lambda j}
\label{subsec:prior-lambda}}

The GP magnitude and length-scale are only weakly identified: similar
data fits can arise from jointly rescaling $\delta_j$ and
$\lambda_j$, inducing strong posterior dependence and poor
MCMC mixing \parencite{paciorek2003}.
When $\lambda_j$ is very small relative to the spacing of the
inputs, the GP shows an undesirable interpolation. 
A prior that penalizes extremely small
$\lambda_j$ avoids this behavior \parencite{Stan}.
For this reason, we place an inverse-gamma prior on the length-scale parameters:
\begin{equation}
\label{eq:prior-lambda}
\lambda_j
\sim
\mathrm{IG}(a=3,\;b=0.8).
\end{equation}
With predictors rescaled so that $\mathrm{range}(x_j)=1$,
this specification has a median close to $0.3$ (a moderate smoothness
level; see Figure~\ref{fig:gp_effects}), concentrates most mass on
$[0,\,0.5]$, and decays rapidly beyond~$1$.
Length-scales much larger than $1$ produce near-constant covariance
across the domain for stationary kernels, so further increases have
negligible effect on the effective smoothness; concentrating little
mass beyond $1$ avoids this uninformative plateau while retaining
flexibility in the relevant range \parencite{Stan}.

\section{Bayes factor computation\label{subsec:bf-computation}}

Bayes factors provide a principled probabilistic measure to quantify the relative evidence between competing hypotheses. Here we explain how to compute Bayes factors for the linear and nonlinear testing problems discussed in Section \ref{sec:testing}.

For a comparison of the full model $M_1$ against a restricted
model $M_0$, the Bayes factor is
\begin{equation}
\label{eq:bf-def}
\mathrm{BF}_{10}
  = \frac{p(\mathbf{y}\mid M_1)}{p(\mathbf{y}\mid M_0)},
\end{equation}
where each marginal likelihood integrates the sampling distribution
over all parameters and hyperparameters under the respective models.
Values for which $\mathrm{BF}_{10}>1$ indicate evidence in favor of $M_1$;
values below $1$ favor $M_0$.

Computing marginal likelihoods involves high-dimensional
integrals over the free parameters that do not admit closed-form
evaluation. This poses a severe computational burden, especially because Gaussian processes are already computationally expensive \parencite{williams2006gaussian, quinoneroCandela2005unifying}.
For the hypothesis tests described in Section \ref{sec:testing}, the marginal likelihoods of $2p+1$ different hypotheses would need to be computed, which limits the use of Bayes factors testing of linear and nonlinear effects for applied research. We can drastically decrease this computational burden by considering a Savage--Dickey-type approach for Bayes factor computation only requiring a single model fit. The computational approach is validated against the computation of the actual marginal likelihoods based on bridge sampling as an independent benchmark. Posterior inference is carried out by Hamiltonian Monte Carlo,
using the no-U-turn sampler as implemented in Stan via the \texttt{rstan}
interface \parencite{Stan}. 
The next subsection describes the computational strategies used
to estimate these quantities.

\subsection{Savage--Dickey density ratio\label{subsubsec:sdr}}

When two models are nested and the tested parameter is a scalar
with a proper prior under the alternative, the Bayes factor can be
computed via the Savage--Dickey density ratio (SDR).
This identity writes $\mathrm{BF}_{10}$ as the ratio of the prior
to the posterior marginal density evaluated at the null value:
\begin{equation}
\label{eq:sdr}
\mathrm{BF}_{10}
  = \frac{\pi(\psi = \psi_0|M_1)}
         {\pi(\psi = \psi_0 \mid \mathbf{y}, M_1)},
\end{equation}
where $\psi$ denotes the parameter being tested and $\psi_0$ its
null value, provided that the prior for the nuisance parameters under the restricted model coincides with their conditional prior under the full model conditional on the tested parameter being zero; prior independence of the tested parameter and the nuisance parameters is a sufficient special case \parencite{Dickey1971,verdinelli1995, mulder2022generalization}.
 
The SDR is attractive in our case because all $2p$ tests use the same full model. Thus, for computing the $2p$ Bayes factors only a single fit of the full model $M_1$ in~\eqref{eq:marginal} is required: the
prior ordinate is available analytically from the priors specified
in Section~\ref{sec:priors}, and the posterior ordinate is
estimated from MCMC output obtained under the fitted full model.

For testing the linear effect $H_0:\theta_j=0$, the required prior
ordinate is obtained from the marginal $g$-prior for $\theta_j$.
Integrating out $g$ under \eqref{eq:prior-g} yields the scaled
$t$-distribution
\begin{equation}
\label{eq:marginal-theta}
\theta_j \;\sim\; t\!\bigl(2a,\; 0,\;
  \sqrt{b\,v_j / a}\bigr),
  \qquad v_j = \bigl[(\mathbf{X}^{\top}\!\mathbf{X})^{-1}\bigr]_{jj},
\end{equation}
which for $a=2$ and $b=n/2$ gives $\theta_j \sim t(4,\,0,\,
\sqrt{n\,v_j/4})$. This marginal prior has a well-defined,
analytically available density at the origin, so the prior ordinate is
computed directly from \eqref{eq:marginal-theta}. The posterior
ordinate is estimated from the MCMC output by logspline density
estimation \parencite{kooperberg1991logspline, WagenmakersEtAl2010},
which limits its accuracy where posterior draws are sparse.
\textcite{MoreyEtAl2011} report that the logspline estimator can be
substantially biased precisely when the Bayes factor is large, and
\textcite{SchadEtAl2022} note that any Savage--Dickey ratio computed
from MCMC output inherits this limitation, because few draws fall near
the null value once the posterior is located far from it. The Bayes
factors reported for strong linear effects in
Section~\ref{sec:empirical} should be read with this in mind. Note that
SDRs can also be used for computing fractional Bayes factors
\parencite{OHagan:1995,mulder2014prior} for testing linear effects as
implemented in the R package BFpack \parencite{mulder2021bfpack}.

For testing the nonlinear effect $H_0:\delta_j=0$, two further
difficulties arise. The first is specific to the geometry of
$\delta_j$: because $\delta_j$ is non-negative, the null value lies at
the boundary of the parameter support rather than in its interior, so
the ordinate has to be read off the edge of the fitted density. Boundary
truncation then adds to the sparse-draw problem described above, and
$\hat\pi(0\mid\mathbf{y})$ is distorted even when the posterior is
moderately concentrated near zero. The second difficulty is that, under
the non-local prior~\eqref{eq:nonlocal-prior} with $d \ge 1$, the prior
density vanishes at the origin by construction, as $\pi_d(0\mid M_1)=0$,
and therefore the marginal posterior density will automatically also be
$\pi_d(0\mid \mathbf{y}, M_1)=0$, so that the SDR becomes ill-defined as
$0/0$. For the nonlinear test, and in particular for the non-local
priors, a different formulation of the Bayes factor identity is
therefore required.

\subsection{Bayes factors for testing nonlinear effects via interval nulls\label{subsubsec:intbf}}

We replace the precise null hypothesis, $H_{0,j}^{(N)}:\delta_j=0$, by an interval null $H_{0,\varepsilon}:\; 0 \le \delta_j \le \varepsilon$. Based on results of encompassing prior approaches to Bayes factor computation \parencite{klugkist2005inequality,mulder2010equality}, the Bayes factor can then be computed as the ratio of prior and posterior probability that the interval constraints hold, i.e.,
\begin{equation}
\label{eq:bf-interval}
\mathrm{BF}_{10}^{(\varepsilon)}
  = \frac{\Pr(0 \le \delta_j \le \varepsilon)}
         {\Pr(0 \le \delta_j \le \varepsilon \mid \mathbf{y})}.
\end{equation}
Using l'H\^opital's rule and the result of \textcite{wetzels2010encompassing}, the Bayes factor approaches the Bayes factor for the precise null test when $\varepsilon$ goes to 0 both for local and non-local priors. A proof is given in Appendix~\ref{app:interval-limit}. Note that the approximation of a precise null by an interval null was also considered by \textcite{berger1987testing}.

Because of this limiting equality, we use \eqref{eq:bf-interval} to approximate the required Bayes factors for testing nonlinear effects. 
As the prior and posterior probabilities are strictly positive, the ill-defined behavior of the SDR for the precise test is avoided. In practice, the interval-null formulation requires the specification of $\varepsilon$. 
The approximated Bayes factor in \eqref{eq:bf-interval} becomes more accurate when $\varepsilon$ is as small as possible.
Estimating the posterior probability of the interval can become numerically inaccurate, however, when few or no posterior draws fall in it.
We therefore estimate this probability by a logspline fit to the posterior draws rather than by their empirical proportion.
On the other hand, obtaining more posterior draws would make the estimator more computationally expensive.
To balance this choice, we take $\varepsilon=0.2$, which corresponds to a relatively small effect as a $\delta$ of 1 can be interpreted as a small nonlinear effect (Figure~\ref{fig:gp_effects}).
The prior probability is available in closed form.
Under the local prior and the three non-local priors, $\Pr(\delta_j\le 0.2)$ equals $0.0549$, $0.0079$, $0.0015$, and $0.0004$ for $d=0,1,2,3$, respectively.
For brevity, we refer to this interval-null implementation as the SDR-based computation in the remainder of the paper.

\subsection{Reweighting Bayes factors for non-local priors\label{subsubsec:reweighting}}

Computing Bayes factors based on the different non-local priors using \eqref{eq:bf-interval} would require refitting the full model for each separate non-local prior. We can avoid this computational burden by expressing the Bayes
factors based on the non-local priors as the Bayes factor based on its base prior and a reweighting factorization which can also be obtained from MCMC output under the base prior.

Let $\pi^{B}(\delta_j)$ denote the base prior for the nonlinear magnitude
parameter and $\pi^{T}(\delta_j)$ the target prior. Under the full model
$M_1$, the marginal likelihood under the target prior is
\begin{equation}
\label{eq:reweighting-marginal}
p^{T}(\mathbf{y}\mid M_1)
  = \int p(\mathbf{y}\mid \delta_j, M_1)\,\pi^{T}(\delta_j)\,d\delta_j.
\end{equation}
Multiplying and dividing the integrand by the base prior
$\pi^{B}(\delta_j)$ leaves the integral unchanged,
\begin{equation}
\label{eq:reweighting-ratio}
p^{T}(\mathbf{y}\mid M_1)
  = \int p(\mathbf{y}\mid \delta_j, M_1)\,\pi^{B}(\delta_j)\,
    \frac{\pi^{T}(\delta_j)}{\pi^{B}(\delta_j)}\,d\delta_j.
\end{equation}
By Bayes' theorem under the base prior,
$p(\mathbf{y}\mid \delta_j, M_1)\,\pi^{B}(\delta_j)
  = p^{B}(\mathbf{y}\mid M_1)\,\pi^{B}(\delta_j\mid \mathbf{y})$,
so the base marginal likelihood factors out of the integral and the
remaining integral is an expectation over the base-prior posterior of
$\delta_j$,
\begin{equation}
\label{eq:reweighting-identity-ml}
p^{T}(\mathbf{y}\mid M_1)
  = p^{B}(\mathbf{y}\mid M_1)
    \int \pi^{B}(\delta_j\mid \mathbf{y})\,
      \frac{\pi^{T}(\delta_j)}{\pi^{B}(\delta_j)}\,d\delta_j
  = p^{B}(\mathbf{y}\mid M_1)\,
    \mathbb{E}^{B}_{\delta_j\mid \mathbf{y}}
    \!\left[\frac{\pi^{T}(\delta_j)}{\pi^{B}(\delta_j)}\right].
\end{equation}
The null model $M_0$ does not involve $\delta_j$ and so is common to both
priors; dividing both sides by $p(\mathbf{y}\mid M_0)$ therefore yields
the Bayes factor identity
\begin{equation}
\label{eq:reweighting-identity-bf}
\mathrm{BF}^{T}_{10}
  = \mathrm{BF}^{B}_{10}\,
    \mathbb{E}^{B}_{\delta_j\mid \mathbf{y}}
    \!\left[\frac{\pi^{T}(\delta_j)}{\pi^{B}(\delta_j)}\right].
\end{equation}
Thus, once $\mathrm{BF}^{B}_{10}$ has been computed under the base prior,
the Bayes factor under any target prior follows by averaging the
prior-density ratio $\pi^{T}(\delta_j)/\pi^{B}(\delta_j)$ over the
posterior draws already available from the base fit, without any
refitting.

We apply this identity with the local ($d=0$) prior as the base prior
$\pi^{B_0}$ and the non-local priors of order $d \in \{1,2,3\}$ as the
targets. Because each order $d$ is calibrated through its own value of
$\kappa$ so that the prior median remains fixed on the standardized
$\delta$ scale, the target and base priors are not related by a simple
multiplicative factor, and we evaluate the density ratio
$\pi_d(\delta_j)/\pi^{B_0}(\delta_j)$ pointwise. Given posterior draws
$\delta_j^{(1)},\dots,\delta_j^{(S)}$ from the base fit, the reweighted
Bayes factor is estimated by
\begin{equation}
\label{eq:reweighting-estimator}
\widehat{\mathrm{BF}}^{\,\pi_d}_{10}
  = \widehat{\mathrm{BF}}^{\,B_0}_{10}\,
    \frac{1}{S}\sum_{s=1}^{S}
    \frac{\pi_d(\delta_j^{(s)})}{\pi^{B_0}(\delta_j^{(s)})}.
\end{equation}
Accordingly, the local-prior Bayes factor is computed directly from the
base fit, and the Bayes factors for the three non-local priors are
obtained from the same fit by reweighting.

\subsection{Bridge sampling as a benchmark\label{subsubsec:bridge}}
The single-fit Bayes factor computations used in
Sections~\ref{subsubsec:sdr} and~\ref{subsubsec:intbf} rely on
density or probability estimation at or near the null value.
As an independent validation check, we
also compute Bayes factors directly using bridge sampling
\parencite{Meng1996, Gronau2017}.

Note, however, that bridge sampling requires a fitting step for every separate model. Thus when testing the linear and nonlinear effects of $p$ predictors against the full model, $2p+1$ models need to be evaluated, and thus the computational time grows linearly with the
number of predictors. The SDR-based testing approach, on the other hand, only requires a single fit of the full model.

\section{Simulation\label{sec:simulation}}

The first goal of the simulation was to examine the behavior of the proposed Bayes factors in a controlled
simulation study by varying the strength of the nonlinear signal, according to $\delta \in \{0,1,2,3\}$, and the sample size, according to $N \in \{20,50,100,200\}$. A second goal was to assess the accuracy of the SDR-based computation of the Bayes factor by comparing it to the bridge sampling estimate. A single predictor was considered which was placed on an evenly spaced grid over $[-0.5,0.5]$. The length-scale was set to $\lambda=0.3$ and the error standard deviation was set to $\sigma=0.1$. For each scenario 50 datasets were generated and Bayes factors were computed based on the four prior orders
$d \in \{0,1,2,3\}$. For the interval-null procedure, the threshold was fixed at $\varepsilon=0.2$ throughout.

All models were fit in Stan through the \texttt{rstan} interface
\parencite{Stan}, running four chains with 500 warmup and 1500
post-warmup iterations each and control settings
$\texttt{adapt\_delta}=0.95$ and $\texttt{max\_treedepth}=12$. Marginal
likelihoods were obtained with the \texttt{bridgesampling} package
\parencite{Gronau2017}, applied separately to the alternative and null
fits. We relied on Stan's standard sampler diagnostics; no divergent
transitions or maximum-tree-depth warnings were raised for the fits
reported here.

\begin{figure*}[htbp]
  \centering
  \includegraphics[width=\textwidth]{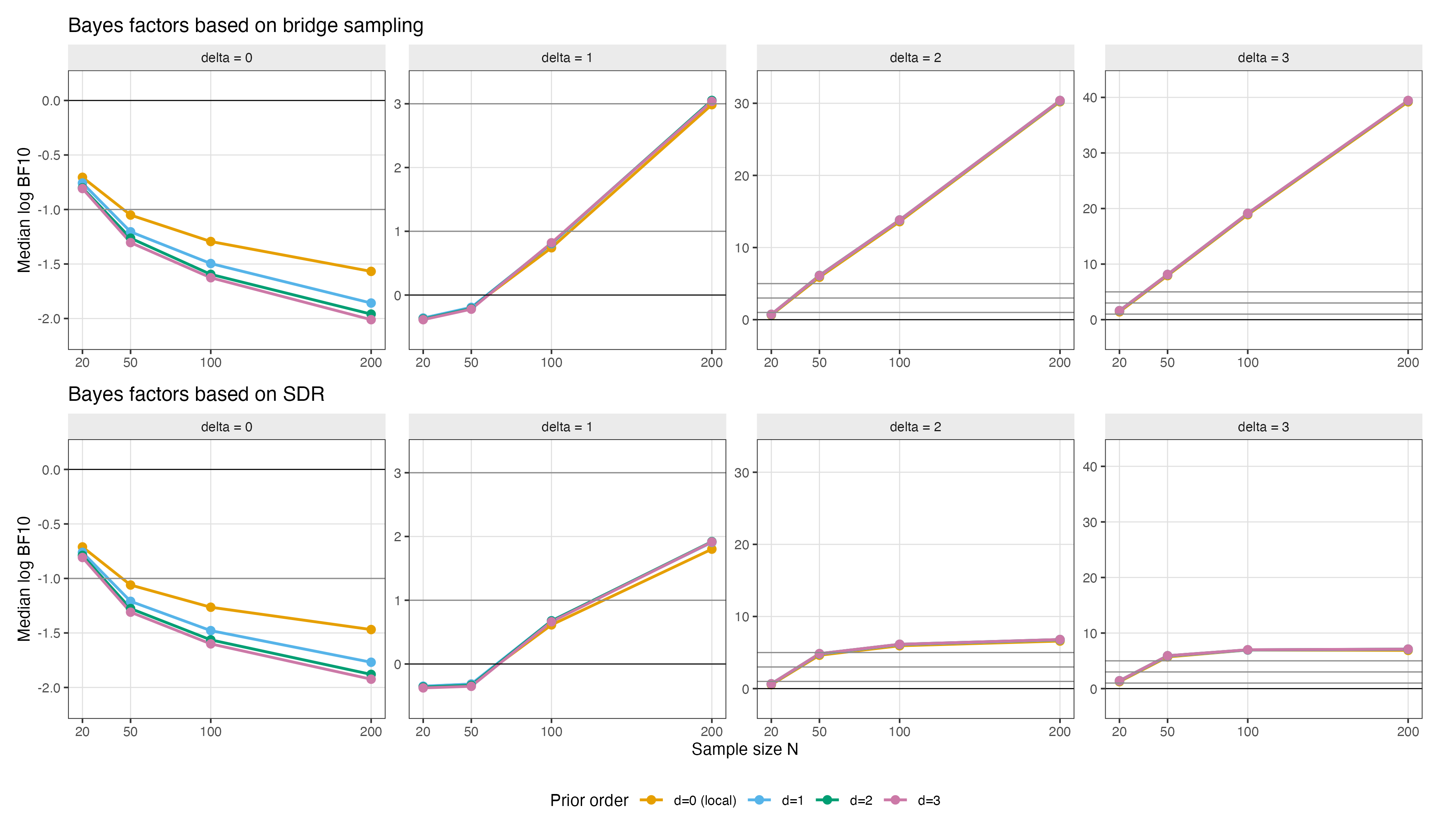}
  \caption{Simulation results for nonlinear-effect Bayes factors under
  $\delta \in \{0,1,2,3\}$, $N \in \{20,50,100,200\}$, fixed
  $\lambda=0.3$, and prior orders $d \in \{0,1,2,3\}$. Columns
  correspond to the true nonlinear magnitude $\delta$, and colors denote
  the prior order. The first row shows direct bridge-sampling Bayes
  factors. The second row shows the SDR-based Bayes factors,
  obtained from the $d=0$ posterior using the interval-null Bayes factor
  with $\varepsilon=0.2$ as the base quantity and reweighting to
  $d=1,2,3$. Points and lines show median log Bayes factors across 50
  simulated datasets per configuration; Monte Carlo variation across
  replicates is not displayed. Note that the vertical scale is free
  across columns, reflecting the rapid growth of the Bayes factor with
  $\delta$. The grey lines depict the evidence thresholds of
  \textcite{kass1995bayes} at $\log\mathrm{BF}=1$ (`positive'), $3$
  (`strong'), and $5$ (`very strong').}
  \label{fig:simulation-bf}
\end{figure*}

The results are summarized in Figure~\ref{fig:simulation-bf}. The first
row gives the bridge-sampling benchmark which shows the anticipated behavior: as the true nonlinear effect increases, the Bayes factor shifts from favoring the null to favoring the alternative, and
this shift becomes more pronounced as the sample size grows. Moreover, the evidence generally points in the correct direction except when the nonlinear effect is small ($\delta=1$) and the sample is relatively small ($N=20$). This can be explained by the fact that relatively large samples are required to identify small nonlinear effects. Finally, we see that the non-local moment priors tend to result in more evidence in favor of the null in case the null is true, as expected. In case the null is false, the priors behave practically identically. The underlying median log Bayes factors are tabulated in Table~\ref{tab:bridge-medians} of Appendix~\ref{app:sim-table}.

The second row reports the SDR-based Bayes factors for testing a nonlinear effect. When the null is true, the general behavior is similar to the bridge-sampling benchmark. When the alternative is true, the two approaches agree closely while the evidence remains below the `very strong' threshold of 5 on the logarithmic scale. When the evidence is very strong (for $\log\mathrm{BF}>5$), the SDR-based computation underestimates the evidence against the null. 
This occurs because, when the posterior places very little mass in the interval $0 \le \delta_j \le \varepsilon$, the logspline estimate of $\Pr(\delta_j\le\varepsilon\mid\mathbf{y})$ has to be extrapolated into a region containing few or no draws.
Nevertheless, the SDR-based computation still clearly favors the correct hypothesis; it is the magnitude, rather than the direction, of the evidence that is affected. Thus, the SDR-based computation is useful for identifying nonlinear effects, although the strength of the evidence may be underestimated when the posterior is far from the null value.

\section{Empirical Applications\label{sec:empirical}}

\subsection{Mother's IQ vs.\ Child Test Scores}

We analyze data from a subsample of the National Longitudinal Survey of Youth, focusing on the relationship between mothers' IQ scores and their children's test scores. The dataset, described by \textcite{gelman2007}, contains $n = 434$ mother--child pairs together with a binary indicator of whether the mother completed high school (HS${}=1$: 341 mothers; HS${}=0$: 93 mothers). This application exercises a group-interaction extension of the framework: in addition to testing whether the average relationship between mother's IQ and child test scores is nonlinear, we test whether the two education groups differ in nonlinear shape after allowing for group differences in their intercepts and linear slopes.

We use deviation coding for the group indicator, $z_i\in\{-\tfrac{1}{2},+\tfrac{1}{2}\}$ for HS${}=0$ and HS${}=1$, respectively, following the modeling of group-specific effects in additive Gaussian process regression \parencite{timonen2021lgpr}. Writing $x$ for mother's IQ, the model is
\begin{equation}
  y_i \;=\; \beta_0 + \beta_1 x_i
         \; + \; \beta_2 z_i + \beta_3 z_i x_i
         \; + \; f_{\text{main}}(x_i)
         \; + \; z_i\,f_{\text{int}}(x_i)
         \; + \; \varepsilon_i,
\end{equation}
\begin{equation}
  f_{\text{main}}\sim \mathcal{GP}\!\bigl(0,\;
      \sigma^2\delta_{\text{main}}^{\,2}\,
      {K}^{\perp}_{\lambda_{\text{main}}}\bigr),
  \qquad
  f_{\text{int}}\sim \mathcal{GP}\!\bigl(0,\;
      \sigma^2\delta_{\text{int}}^{\,2}\,
      {K}^{\perp}_{\lambda_{\text{int}}}\bigr).
\end{equation}
With this coding, the average of the two group-specific trends is represented by $\beta_0+\beta_1x+f_{\text{main}}(x)$, whereas their difference is represented by $\beta_2+\beta_3x+f_{\text{int}}(x)$. Both GP components are orthogonalized with respect to $\operatorname{span}\{1,x\}$, following Section~\ref{subsec:orthogonalization}. Consequently, the Bayes factor for $\delta_{\text{main}}=0$ tests whether the average relationship requires nonlinear structure beyond a linear trend, whereas the Bayes factor for $\delta_{\text{int}}=0$ tests whether the groups differ in nonlinear shape after allowing for differences in their intercepts and linear slopes. The local prior ($d=0$) is applied to both $\delta_{\text{main}}$ and $\delta_{\text{int}}$.

\begin{figure}[htbp]
  \centering
  \includegraphics[width=\textwidth]{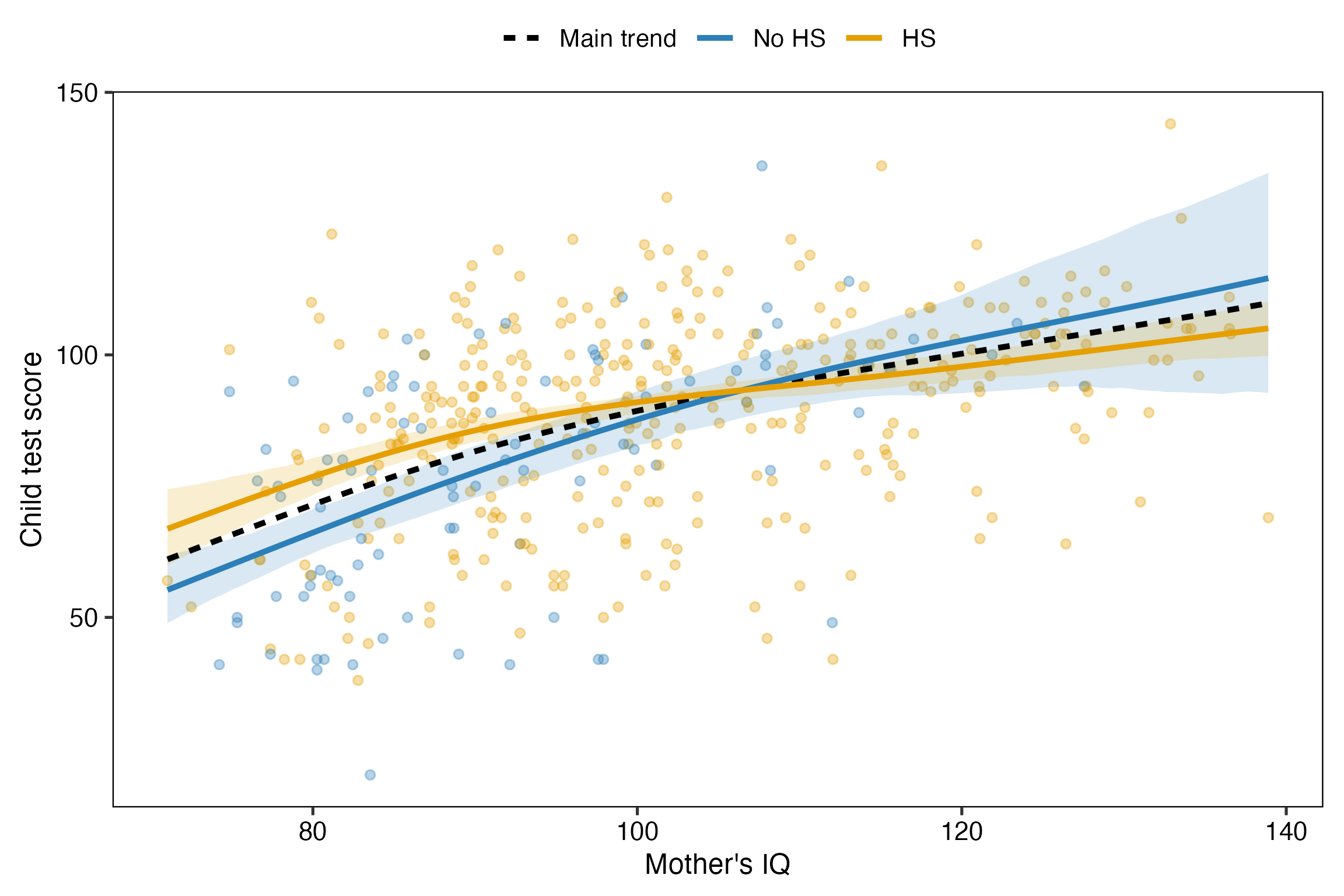}
  \caption{Mother's IQ (horizontal) vs.\ child test scores (vertical). The dashed black curve shows the estimated average trend across the two education groups; solid curves show the group-specific fitted trends for mothers without high school completion (blue) and with high school completion (orange), with shaded regions indicating $95\%$ credible intervals. Points show the observed data for the two groups.}
  \label{fig:education}
\end{figure}

The results are reported in Table~\ref{tab:bf_education}.
There is very strong evidence for a positive linear association between mother's IQ and child test scores. In contrast, there is evidence against an intercept difference between the two education groups, and there is no clear evidence for a linear interaction between IQ and education group. Neither the nonlinear main effect nor the nonlinear interaction receives evidence in favor of the alternative.
Thus, after allowing the groups to differ in their intercepts and linear slopes, the data provide no evidence that either the average IQ--test-score relationship or the difference between the two groups requires additional nonlinear structure.
The two computations agree to within $0.3$ log units for the remaining effects, whereas for the linear main effect they differ by more than four, which reflects the limitation noted in Section~\ref{subsubsec:sdr}: the posterior lies far from zero, where the logspline ordinate is less accurate.
Figure~\ref{fig:education} shows the fitted average and group-specific trends. The marginal prior and posterior distributions of all five tested parameters are shown in Appendix~\ref{app:posteriors} (Figure~\ref{fig:post-a2}).

\begin{table}[t]
  \centering
  \caption{Log Bayes factors ($\log\mathrm{BF}_{10}$) for the linear and nonlinear effects in the mother's IQ model. The linear effects are tested with the Savage--Dickey density ratio; for the nonlinear effects the SDR column uses the $\varepsilon=0.2$ interval null under the local prior ($d=0$). BS: bridge sampling.}
  \label{tab:bf_education}
  \begin{tabular}{llcc}
    \toprule
    Component & Parameter & $\log\mathrm{BF}_{10}^{\text{SDR}}$ & $\log\mathrm{BF}_{10}^{\text{BS}}$ \\
    \midrule
    Main (linear)           & $\beta_1$                 & 13.25 & 17.62 \\
    Intercept difference    & $\beta_2$                 & -2.29 & -2.29 \\
    Interaction (linear)    & $\beta_3$                 & -0.46 & -0.45 \\
    Main (nonlinear)        & $\delta_{\text{main}}$    & -0.49 & -0.24 \\
    Interaction (nonlinear) & $\delta_{\text{int}}$     & -0.92 & -1.01 \\
    \bottomrule
  \end{tabular}
\end{table}

\subsection{Diabetes Dataset}

The diabetes dataset of \textcite{efron2004} records disease progression for $n=442$ patients. We fit the additive GP model to the nine continuous predictors: \texttt{age}, \texttt{bmi}, \texttt{bp} (mean arterial pressure), and six blood serum measurements, corresponding to total cholesterol (\texttt{s1}), LDL (\texttt{s2}), HDL (\texttt{s3}), total cholesterol/HDL (\texttt{s4}), log triglycerides (\texttt{s5}), and glucose (\texttt{s6}).\footnote{This naming convention is standard for the diabetes benchmark dataset.} In benchmark analyses, \textcite{efron2004} used this dataset to illustrate the LARS/LASSO solution paths, and \textcite{park2008bayesian} applied the Bayesian Lasso; both studies highlight that only a small subset of predictors carries most of the predictive signal, with \texttt{bmi}, \texttt{bp}, and \texttt{s5} repeatedly among the most important. This earlier literature provides a useful benchmark for our decomposition, because it identifies influential predictors but does not distinguish whether their contribution is primarily linear or nonlinear.

\begin{table}[t]
  \centering
  \caption{Predictor-specific evidence in the Diabetes dataset. The Bayes-factor columns report $\log(\mathrm{BF}_{10})$ for the linear and nonlinear effects under the local prior ($d=0$); the nonlinear Bayes factors under the non-local priors ($d=1,2,3$) are reported in Table~\ref{tab:reweight_diabetes} of Appendix~\ref{app:reweight}. The ARD column reports the inverse range from the anisotropic GP baseline, so larger values indicate shorter length-scales and greater overall GP sensitivity.}
  \label{tab:diabetes_bf_ard}
  \small
  \begin{tabular}{lrrr}
    \toprule
    Predictor & Linear & Nonlinear & ARD \\
    \midrule
    age & -2.26 & -0.17 & 0.139 \\
    bmi & 19.09 & -0.85 & 0.173 \\
    bp  & 5.85  & -1.24 & 0.107 \\
    s1  & -0.94 & -1.43 & 0.050 \\
    s2  & -1.66 & -1.17 & 0.000 \\
    s3  & -2.19 & -1.46 & 0.055 \\
    s4  & -2.12 & -1.39 & 0.000 \\
    s5  & 5.37  & -0.12 & 0.262 \\
    s6  & -2.14 & -0.06 & 0.083 \\
    \bottomrule
  \end{tabular}
\end{table}

\begin{figure}[t]
  \centering
  \includegraphics[width=\linewidth]{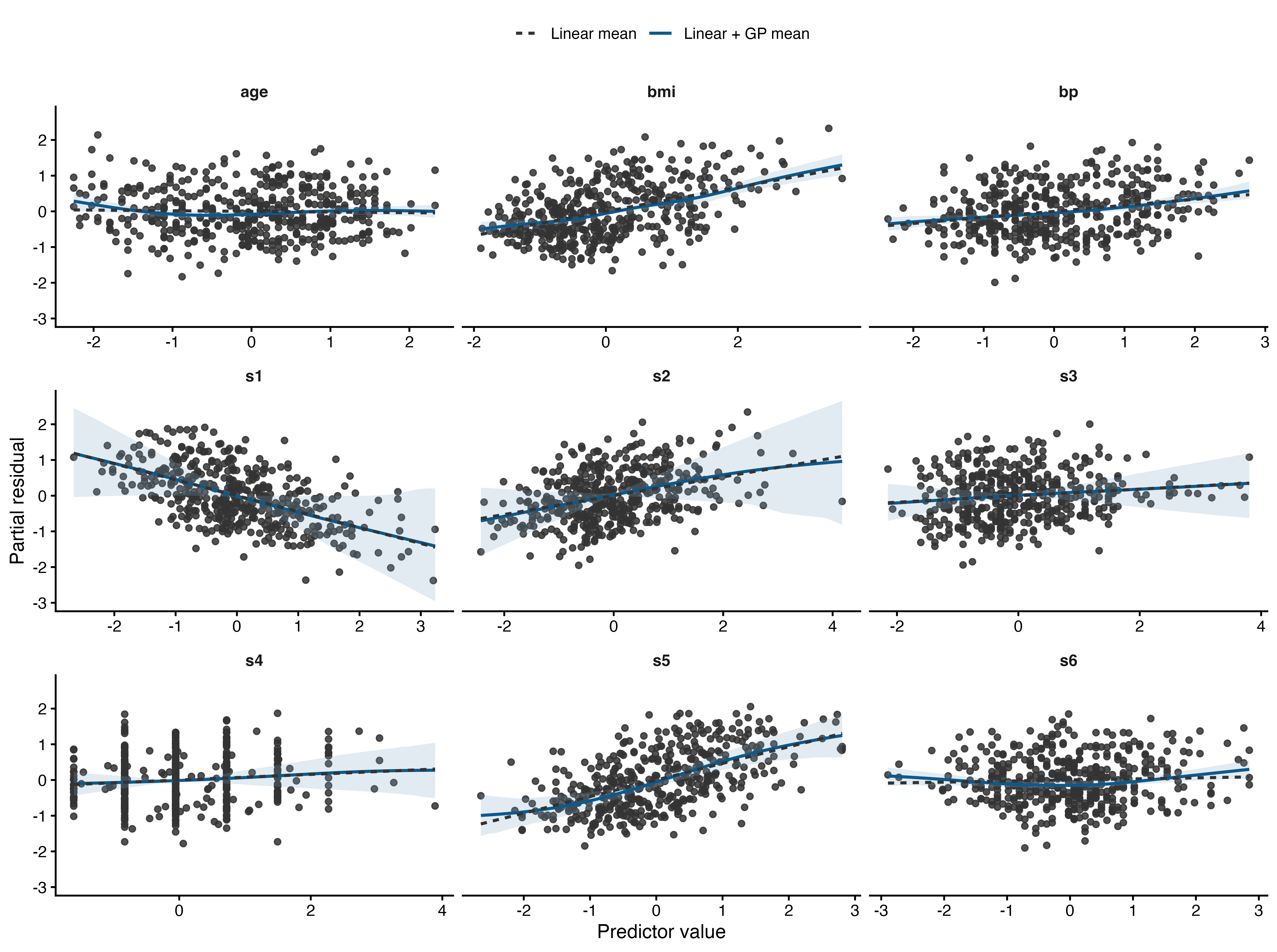}
  \caption{Estimated predictor effects in the Diabetes dataset under the local prior ($d=0$). Each panel shows the partial effect of one predictor: grey points are partial residuals, i.e.\ the standardized response with the fitted contributions of all other predictors removed, so that the scatter reflects only the variation attributable to that predictor. The solid blue curve is the posterior mean of the predictor's additive component (linear plus Gaussian process) with its pointwise $95\%$ credible band, and the dashed grey line is the posterior mean of its linear component alone. The near coincidence of the two curves in every panel is consistent with the absence of nonlinear evidence in Table~\ref{tab:diabetes_bf_ard}.}
  \label{fig:diabetes_trend}
\end{figure}

Table~\ref{tab:diabetes_bf_ard} shows that the diabetes signal is predominantly linear. Regarding the evidence for linear effects, strong to very strong evidence is obtained for the linear effects of \texttt{bmi}, \texttt{s5}, and \texttt{bp}, whereas the remaining predictors have negative log Bayes factors for the linear component. This agrees closely with the classical importance pattern reported by \textcite{efron2004} and \textcite{park2008bayesian}, and sharpens it by showing that the support for these predictors is carried mainly through linear effects rather than through nonlinear departures. For the nonlinear components, all predictors have negative log Bayes factors, which are based on the local prior. Table~\ref{tab:reweight_diabetes} in Appendix~\ref{app:reweight} reports the Bayes factors based on the non-local prior which yield equivalent conclusions. Thus, we find no evidence in favor of nonlinear structure for any of the predictors. The estimated predictor effects in Figure~\ref{fig:diabetes_trend} confirm this visually: in every panel the additive fit (solid) stays close to its linear component (dashed). The full marginal posteriors of the linear and nonlinear parameters for this application are shown in Appendix~\ref{app:posteriors} (Figures~\ref{fig:post-a3-delta} and \ref{fig:post-a3-theta}).

The ARD comparison is qualitatively consistent with this interpretation, but it is less specific. Predictors such as \texttt{s5}, \texttt{bmi}, and \texttt{bp} receive relatively short ARD ranges, which agrees with their overall importance, yet ARD does not reveal that this importance is primarily linear. The contrast is most visible for \texttt{age} and \texttt{s6}: both receive moderate ARD scores despite weak evidence for either a linear or a nonlinear effect in our decomposition. At the other extreme, \texttt{s2} and \texttt{s4} receive ARD values of essentially zero, indicating that the anisotropic GP effectively switches these predictors off. This illustrates the role of ARD in our study: it provides a coarse GP-based sensitivity summary, whereas the proposed Bayes-factor analysis separates evidence for linear and nonlinear contributions.

\subsection{Boston Housing Dataset}

The Boston Housing dataset, originally compiled by \textcite{harrison1978}, contains data on 506 census tracts in the Boston area and has become a standard benchmark for regression and variable selection methods. In our analysis, we exclude the binary predictor \texttt{chas} and the ordinal accessibility measure \texttt{rad}, and retain the 11 continuous predictors \texttt{crim}, \texttt{zn}, \texttt{indus}, \texttt{nox}, \texttt{rm}, \texttt{age}, \texttt{dis}, \texttt{tax}, \texttt{ptratio}, \texttt{black}, and \texttt{lstat}.\footnote{The \texttt{black} predictor is a transformed measure of racial composition, $1000(\mathrm{Bk}-0.63)^2$, whose construction and ethical implications have been widely criticized; it has since been removed from several standard distributions of the dataset.} We retain these only for comparability with earlier analyses. The response variable is the median value of owner-occupied homes (in thousands of dollars). This dataset has been analyzed repeatedly as a variable-selection benchmark, and GP-based analyses in particular often emphasize predictors such as \texttt{nox}, \texttt{rm}, \texttt{dis}, \texttt{tax}, and \texttt{lstat}. In a directly relevant comparison, \textcite{savitsky2011} used a fully multivariate GP with spike-and-slab selection on kernel parameters and also highlighted these variables as important. These earlier findings provide a natural reference point for our results.

\begin{table}[t]
  \centering
  \caption{Predictor-specific evidence in the Boston Housing dataset. The Bayes-factor columns report $\log(\mathrm{BF}_{10})$ for the linear and nonlinear effects under the local prior ($d=0$); the nonlinear Bayes factors under the non-local priors ($d=1,2,3$) are reported in Table~\ref{tab:reweight_boston} of Appendix~\ref{app:reweight}. The ARD column reports the inverse range from the anisotropic GP baseline, so larger values indicate shorter length-scales and greater overall GP sensitivity.}
  \label{tab:boston_bf_ard}
  \small
  \begin{tabular}{lrrr}
    \toprule
    Predictor & Linear & Nonlinear & ARD \\
    \midrule
    crim    & 2.52  & 3.22  & 0.093 \\
    zn      & 6.49  & -1.30 & 0.003 \\
    indus   & -1.89 & 3.51  & 0.133 \\
    nox     & 9.78  & 10.45 & 0.450 \\
    rm      & 31.63 & 11.33 & 0.243 \\
    age     & -2.47 & -1.65 & 0.172 \\
    dis     & 28.02 & 8.76  & 0.792 \\
    tax     & -2.33 & 6.97  & 1.054 \\
    ptratio & 18.81 & -1.31 & 0.098 \\
    black   & 6.58  & -0.91 & 0.312 \\
    lstat   & 34.81 & 6.82  & 0.566 \\
    \bottomrule
  \end{tabular}
\end{table}

\begin{figure}[t]
  \centering
  \includegraphics[width=\linewidth]{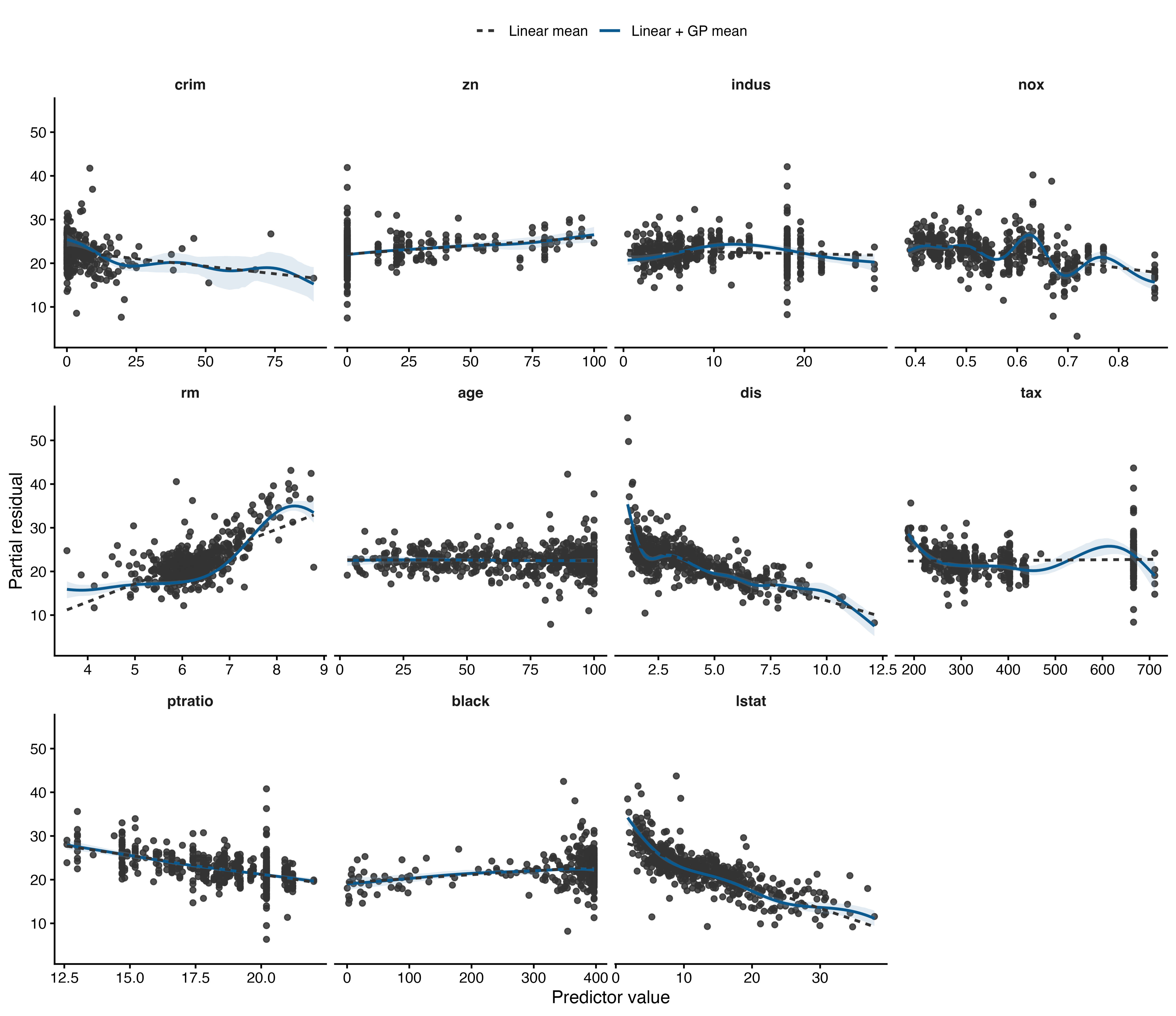}
  \caption{Estimated predictor effects in the Boston Housing dataset under the local prior ($d=0$). Each panel shows the partial effect of one predictor: grey points are partial residuals, i.e.\ the median home value (in \$1000s) with the fitted contributions of all other predictors removed, so that the scatter reflects only the variation attributable to that predictor. The solid blue curve is the posterior mean of the predictor's additive component (linear plus Gaussian process) with its pointwise $95\%$ credible band, and the dashed grey line is the posterior mean of its linear component alone. Predictors with strong nonlinear evidence in Table~\ref{tab:boston_bf_ard}---notably \texttt{rm}, \texttt{lstat}, \texttt{nox}, \texttt{dis}, and \texttt{tax}---show visible departures of the solid curve from the dashed line, whereas \texttt{zn}, \texttt{age}, \texttt{ptratio}, and \texttt{black} remain close to linear.}  \label{fig:boston_trend}
\end{figure}

The Bayes factors in Table~\ref{tab:boston_bf_ard} suggest a more complex dependency pattern than the diabetes application. Several predictors display strong linear evidence, especially \texttt{rm}, \texttt{lstat}, \texttt{dis}, \texttt{ptratio}, and \texttt{nox}, with additional positive support for \texttt{black}, \texttt{zn}, and \texttt{crim}. At the same time, the nonlinear analysis identifies clear departures from linearity for \texttt{nox}, \texttt{rm}, \texttt{dis}, \texttt{tax}, and \texttt{lstat}, while \texttt{crim} and \texttt{indus} show weaker but still positive nonlinear support. This is broadly consistent with previous work, especially the emphasis on \texttt{nox}, \texttt{rm}, \texttt{dis}, \texttt{tax}, and \texttt{lstat}, but the present decomposition is more informative. In particular, it suggests that \texttt{tax} is supported mainly through a nonlinear component, whereas \texttt{ptratio}, \texttt{zn}, and \texttt{black} are supported through linear effects only. These patterns are visible in the estimated effects of Figure~\ref{fig:boston_trend}, where the predictors with nonlinear support show clear curvature while the linear-only predictors remain close to a straight line. The corresponding marginal posteriors are reported in Appendix~\ref{app:posteriors} (Figures \ref{fig:post-a4-delta} and \ref{fig:post-a4-theta}).

The ARD results show clearer qualitative agreement with our nonlinear findings here than in the diabetes application. Predictors with strong nonlinear evidence under our method—especially \texttt{tax}, \texttt{dis}, \texttt{lstat}, \texttt{nox}, and \texttt{rm}—also tend to have relatively short ARD ranges. The agreement is nevertheless not exact. The variable \texttt{black} is a useful counterexample: it receives a comparatively large ARD score, yet our analysis yields weak nonlinear evidence and stronger support for a linear effect. This again underscores the distinction between the two approaches. ARD summarizes overall sensitivity of the fitted GP surface, whereas the proposed Bayes factor analysis separates that sensitivity into evidence for linear and nonlinear structure. Moreover, the proposed Bayes factors provide a clear scale of the relative evidence across models (with a clear neutral quantification around 0 on the log scale), unlike the ARD.

\section{Discussion\label{sec:discussion}}
We have proposed a Bayesian framework for quantifying the evidence for
linear and nonlinear effects in additive Gaussian process regression.
The starting point is an orthogonalized GP construction that separates
each predictor's linear trend from its nonlinear departure. By parameterizing the $p$ nonlinear effects and the $p$ linear effects on a standardized scale, prior specification of these key parameters is simplified. For the standardized nonlinear effect, a half $t$ prior is used as well as non-local extensions. Mixtures of $g$ priors were used for the linear effects. Using Savage--Dickey-type identities, marginal likelihood computation of all $2p+1$ models can be avoided, which is particularly important as GP modeling scales cubically with the sample size. Instead only a single fit of the full model is required. For the two largest applications, this single fit required approximately 4 hours for the diabetes data ($N=442$, $p=9$) and 16 hours for the Boston housing data ($N=506$, $p=11$), running the four MCMC chains in parallel. Obtaining the same linear and nonlinear Bayes factors through marginal likelihoods would instead require a separate fit for each of the $2p+1$ hypotheses, about 19 and 23 fits respectively, on the order of several days and two weeks of computation. This quickly becomes impractical as the number of predictors grows. Simulations revealed that the outcomes are accurate in comparison to Bayes factors using the actual marginal likelihoods, except when the evidence in favor of a nonlinear effect is very strong. In this case, the SDR-based Bayes factors underestimated the magnitude of the evidence while preserving its direction. The SDR-based computation can be viewed as a practical way for testing linear and nonlinear effects in nonlinear GP models. The empirical applications showed how the proposed methods can be used in practice to identify linear and nonlinear effects.

Throughout the paper
we used
exact GP inference. Integrating sparse pseudo-input methods
\parencite{snelson2006sparse, quinoneroCandela2005unifying},
variational inducing-point schemes \parencite{titsias2009variational,
hensman2013gaussian}, or Hilbert space approximations
\parencite{riutortMayol2023hsgp} would improve scalability. The exploration of these approaches falls outside the scope of the current paper but represents a useful direction for future work.

The present
framework treated nonlinear contributions additively, with one component
per predictor. Replacing each univariate component $f_j(x_j)$ with a
component $f_S(\mathbf{x}_S)$ indexed by a subset
$S \subseteq \{1,\dots,p\}$ yields a functional ANOVA-style decomposition
in which each component carries its own standardized magnitude~$\delta_S$
and its own orthogonalized kernel, with higher-order components
orthogonalized against lower-order ones to retain identifiability. The same Bayes factor
machinery can naturally be used for testing such nonlinear interaction structure, see also \textcite{duvenaud2013structure} for example.

The framework is also not tied to a particular covariance kernel. We
used the squared-exponential kernel, but the infinite
differentiability of its sample paths imposes more smoothness than certain
applied settings may warrant \parencite{stein1999interpolation}. The Mat\'ern
family provides a common, less smooth alternative that indexes sample-path
regularity through a parameter~$\nu > 0$:
\begin{equation}
k_{\nu}(x, x' \mid \lambda_j)
  = \frac{2^{1-\nu}}{\Gamma(\nu)}
    \!\left(\frac{\sqrt{2\nu}\,|x-x'|}{\lambda_j}\right)^{\!\nu}
    K_{\nu}\!\left(\frac{\sqrt{2\nu}\,|x-x'|}{\lambda_j}\right),
\end{equation}
where $K_{\nu}$ is the modified Bessel function of the second kind;
common choices are $\nu=3/2$ and $\nu=5/2$, with $\nu\to\infty$
recovering the squared-exponential kernel. Utilizing the proposed standardized parameterization,
the covariance of the nonlinear Mat\'ern component could be written as
$\sigma^2\delta_j^{\,2}\mathbf{K}_{\nu,j}$, retaining its standardized signal-to-noise interpretation of $\delta_j$.
The Bayes factor machinery thus transfers without modification, and
systematic comparison of kernels is a natural follow-up.

The proposed framework is somewhat comparable to regularization-type methods that identify which predictors have an effect and which do not. An alternative approach would be to explore the full model space as in Bayesian variable selection problems. For the current problem, this would imply exploring all possible combinations of predictors having no effect, only a linear effect, only a nonlinear effect, or both a linear and nonlinear effect, yielding $4^p$ possible models. In this situation, inclusion probabilities need to be modeled to handle the multiplicity problem \parencite{scott2010,gu2022}. The computational burden of such model search algorithms will be considerably more demanding, however. Still, the proposed approach would also be computationally beneficial in such algorithms. We leave this for future research.

\appendix
\titleformat{\section}{\normalfont\large\bfseries}{Appendix~\thesection}{1em}{}

\section{Limit of the interval-null Bayes factor}
\label{app:interval-limit}
 
We show that
$\lim_{\varepsilon\to 0}
  \mathrm{BF}_{10}^{(\varepsilon)} = \mathrm{BF}_{10}$
for the moment prior $\pi_d(\delta)=\delta^d\,g(\delta)/C_d$
of order $d\ge 0$.
 
Write $N(\varepsilon)$ and $D(\varepsilon)$ for the numerator and
denominator of~\eqref{eq:bf-interval}:
\[
N(\varepsilon) = \int_0^{\varepsilon}\pi_d(\delta)\,d\delta,
\qquad
D(\varepsilon) = \int_0^{\varepsilon}\pi_d(\delta\mid\mathbf{y})
  \,d\delta.
\]
Both are continuous, vanish at $\varepsilon=0$, and are
differentiable for $\varepsilon>0$ by the fundamental theorem of
calculus, with
\[
N'(\varepsilon) = \pi_d(\varepsilon), \qquad
D'(\varepsilon) = \pi_d(\varepsilon\mid\mathbf{y}).
\]
 
\paragraph{Case $d=0$ (local prior).}
Here $\pi_0(0)=g(0)/C_0>0$ and, by assumption, the conditional
marginal likelihood $p(\mathbf{y}\mid\delta)$ is continuous and
positive at $\delta=0$, so
$\pi_0(0\mid\mathbf{y})>0$.
Since $N(0)=D(0)=0$ and $D'(0)=\pi_0(0\mid\mathbf{y})>0$,
l'H\^opital's rule gives
\[
\lim_{\varepsilon\to 0}
  \frac{N(\varepsilon)}{D(\varepsilon)}
= \frac{N'(0)}{D'(0)}
= \frac{\pi_0(0)}{\pi_0(0\mid\mathbf{y})},
\]
which is the Savage--Dickey density ratio and equals
$\mathrm{BF}_{10}$ by the identity of
\textcite{Dickey1971, verdinelli1995}.
 
\paragraph{Case $d\ge 1$ (non-local prior).}
Both $\pi_d(\delta)$ and $\pi_d(\delta\mid\mathbf{y})$ vanish at
$\delta=0$, so a single application of l'H\^opital's rule yields
$0/0$ again.
We proceed by characterizing the local behavior of both densities
near the origin.
 
Because $g$ is continuous with $g(0)>0$, the prior satisfies
\begin{equation}
\label{eq:prior-local-behavior}
\pi_d(\delta) = \frac{\delta^d\,g(\delta)}{C_d}
  \;\sim\; \frac{g(0)}{C_d}\,\delta^d
  \qquad\text{as } \delta\to 0.
\end{equation}
For the posterior, Bayes' theorem gives
\[
\pi_d(\delta\mid\mathbf{y})
  = \frac{p(\mathbf{y}\mid\delta)\,\pi_d(\delta)}
         {p(\mathbf{y}\mid M_1)},
\]
where $p(\mathbf{y}\mid\delta)
  = \int p(\mathbf{y}\mid\delta,\boldsymbol{\eta})\,
    \pi(\boldsymbol{\eta})\,d\boldsymbol{\eta}$
is the conditional marginal likelihood, integrating over the
nuisance parameters $\boldsymbol{\eta}=(\boldsymbol{\theta},
\boldsymbol{\lambda},\sigma^2)$.
By assumption, $p(\mathbf{y}\mid\delta)$ is continuous and positive
at $\delta=0$; in particular,
$p(\mathbf{y}\mid\delta=0)=p(\mathbf{y}\mid M_0)>0$.
Hence
\begin{equation}
\label{eq:post-local-behavior}
\pi_d(\delta\mid\mathbf{y})
  \;\sim\; \frac{p(\mathbf{y}\mid M_0)}{p(\mathbf{y}\mid M_1)}
    \cdot \frac{g(0)}{C_d}\,\delta^d
  \qquad\text{as } \delta\to 0.
\end{equation}
Both densities vanish as $\delta^d$, so their integrals over
$(0,\varepsilon)$ vanish as $\varepsilon^{d+1}$:
\[
N(\varepsilon) \;\sim\; \frac{g(0)}{C_d(d+1)}\,\varepsilon^{d+1},
\qquad
D(\varepsilon) \;\sim\;
  \frac{p(\mathbf{y}\mid M_0)}{p(\mathbf{y}\mid M_1)}
  \cdot \frac{g(0)}{C_d(d+1)}\,\varepsilon^{d+1}.
\]
The numerator and denominator vanish at the same rate, so the
ratio converges to a finite, positive limit
\parencite[cf.][Section~2]{wetzels2010encompassing}:
\begin{equation}
\label{eq:interval-limit-proof}
\lim_{\varepsilon\to 0}
  \frac{N(\varepsilon)}{D(\varepsilon)}
= \frac{g(0)/\bigl[C_d(d+1)\bigr]}
       {\dfrac{p(\mathbf{y}\mid M_0)}{p(\mathbf{y}\mid M_1)}
         \cdot \dfrac{g(0)}{C_d(d+1)}}
= \frac{p(\mathbf{y}\mid M_1)}{p(\mathbf{y}\mid M_0)}
= \mathrm{BF}_{10}.
\end{equation}
All factors involving $g(0)$, $C_d$, and $(d+1)$ cancel,
confirming that the interval Bayes factor recovers the exact
point-null Bayes factor for any prior order $d\ge 0$.

\section{Simulation Bayes factors}
\label{app:sim-table}

Table~\ref{tab:bridge-medians} reports the median bridge-sampling log Bayes
factors that underlie the first row of Figure~\ref{fig:simulation-bf}, for every
combination of the true nonlinear magnitude $\delta$, the sample size $N$, and the
prior order $d$.

\begin{table}[htbp]
  \centering
  \caption{Median bridge-sampling log Bayes factors
  ($\log \mathrm{BF}_{10}$) across 50 simulated datasets, by true
  nonlinear magnitude $\delta$, sample size $N$, and prior order
  $d$. When $\delta=0$, the non-local priors yield more evidence in
  favor of the null as the order increases. When $\delta>0$, the four
  prior orders behave practically identically: across all sample sizes
  the spread over $d$ does not exceed $0.3$ on the log scale, which is
  negligible relative to the magnitude of the evidence itself.}
  \label{tab:bridge-medians}
  \begin{tabular}{ccrrrr}
    \toprule
    & & \multicolumn{4}{c}{Prior order $d$} \\
    \cmidrule(lr){3-6}
    $\delta$ & $N$ & $d=0$ & $d=1$ & $d=2$ & $d=3$ \\
    \midrule
    0 & 20 & -0.66 & -0.69 & -0.72 & -0.75 \\
     & 50 & -1.07 & -1.23 & -1.27 & -1.31 \\
     & 100 & -1.25 & -1.45 & -1.56 & -1.60 \\
     & 200 & -1.57 & -1.86 & -1.96 & -1.99 \\
    \midrule
    1 & 20 & -0.93 & -1.04 & -1.09 & -1.11 \\
     & 50 & 0.42 & 0.47 & 0.47 & 0.46 \\
     & 100 & 4.85 & 5.00 & 5.01 & 5.01 \\
     & 200 & 7.36 & 7.49 & 7.51 & 7.51 \\
    \midrule
    2 & 20 & 0.12 & 0.19 & 0.20 & 0.19 \\
     & 50 & 6.72 & 6.91 & 6.96 & 6.97 \\
     & 100 & 19.00 & 19.20 & 19.30 & 19.30 \\
     & 200 & 33.50 & 33.70 & 33.80 & 33.80 \\
    \midrule
    3 & 20 & 1.67 & 1.82 & 1.86 & 1.86 \\
     & 50 & 13.20 & 13.40 & 13.40 & 13.40 \\
     & 100 & 32.00 & 32.20 & 32.30 & 32.30 \\
     & 200 & 59.40 & 59.60 & 59.70 & 59.70 \\
    \bottomrule
  \end{tabular}
\end{table}

\section{Nonlinear Bayes factors under non-local priors}
\label{app:reweight}

Tables~\ref{tab:reweight_diabetes} and~\ref{tab:reweight_boston} report the
nonlinear-effect Bayes factors for the two higher-dimensional applications
under the local prior ($d=0$, as reported in the main text) and under the
non-local moment priors of order $d=1,2,3$. As described in
Section~\ref{subsubsec:reweighting}, the non-local values are not obtained by
refitting: each is computed from the single $d=0$ fit through the importance
reweighting identity, so all four columns are read off the same posterior
draws. The $d=0$ columns therefore reproduce the nonlinear columns of
Tables~\ref{tab:diabetes_bf_ard} and~\ref{tab:boston_bf_ard} exactly.

The two applications illustrate the two directions in which a non-local prior
can move the evidence. For the diabetes predictors, where the nonlinear log Bayes factors are already negative at $d=0$, increasing the prior order pushes them further below zero, strengthening the case against a nonlinear
component. For Boston Housing the effect is discriminative: predictors with
genuine nonlinear structure (\texttt{nox}, \texttt{rm}, \texttt{dis},
\texttt{tax}, \texttt{lstat}, and to a lesser degree \texttt{crim} and
\texttt{indus}) gain evidence as $d$ grows, whereas the near-linear predictors
(\texttt{zn}, \texttt{age}, \texttt{ptratio}, \texttt{black}) lose it. In both
cases the qualitative conclusions are stable across all four prior orders.

\begin{table}[t]
  \centering
  \caption{Diabetes dataset: nonlinear-effect log Bayes factors $\log(\mathrm{BF}_{10})$ across moment-prior orders. The $d=0$ column reproduces the nonlinear column of Table~\ref{tab:diabetes_bf_ard}; the $d=1,2,3$ columns are obtained by reweighting the same $d=0$ fit.}
  \label{tab:reweight_diabetes}
  \small
  \begin{tabular}{lrrrr}
    \toprule
    Predictor & $d=0$ & $d=1$ & $d=2$ & $d=3$ \\
    \midrule
    age     &  -0.17 &  -0.19 &  -0.23 &  -0.25 \\
    bmi     &  -0.85 &  -0.99 &  -1.07 &  -1.10 \\
    bp      &  -1.24 &  -1.43 &  -1.51 &  -1.54 \\
    s1      &  -1.43 &  -1.68 &  -1.77 &  -1.81 \\
    s2      &  -1.17 &  -1.34 &  -1.40 &  -1.43 \\
    s3      &  -1.46 &  -1.73 &  -1.83 &  -1.87 \\
    s4      &  -1.39 &  -1.64 &  -1.74 &  -1.77 \\
    s5      &  -0.12 &  -0.11 &  -0.12 &  -0.13 \\
    s6      &  -0.06 &  -0.08 &  -0.11 &  -0.13 \\
    \bottomrule
  \end{tabular}
\end{table}

\begin{table}[t]
  \centering
  \caption{Boston Housing dataset: nonlinear-effect log Bayes factors $\log(\mathrm{BF}_{10})$ across moment-prior orders. The $d=0$ column reproduces the nonlinear column of Table~\ref{tab:boston_bf_ard}; the $d=1,2,3$ columns are obtained by reweighting the same $d=0$ fit.}
  \label{tab:reweight_boston}
  \small
  \begin{tabular}{lrrrr}
    \toprule
    Predictor & $d=0$ & $d=1$ & $d=2$ & $d=3$ \\
    \midrule
    crim    &   3.22 &   3.40 &   3.44 &   3.44 \\
    zn      &  -1.30 &  -1.53 &  -1.62 &  -1.66 \\
    indus   &   3.51 &   3.66 &   3.69 &   3.69 \\
    nox     &  10.45 &  10.72 &  10.80 &  10.82 \\
    rm      &  11.33 &  11.49 &  11.54 &  11.55 \\
    age     &  -1.65 &  -2.01 &  -2.13 &  -2.17 \\
    dis     &   8.76 &   8.77 &   8.78 &   8.79 \\
    tax     &   6.97 &   7.09 &   7.13 &   7.13 \\
    ptratio &  -1.31 &  -1.54 &  -1.63 &  -1.67 \\
    black   &  -0.91 &  -1.03 &  -1.08 &  -1.11 \\
    lstat   &   6.82 &   6.96 &   7.01 &   7.01 \\
    \bottomrule
  \end{tabular}
\end{table}

\section{Posterior samples for the empirical applications}
\label{app:posteriors}

For the three applications of Section~\ref{sec:empirical} we report the
marginal posterior distributions that underlie the Bayes factors
summarized in the main text. In each panel the grey histogram is the
posterior sample and the red curve is the corresponding prior density;
in the panels for the linear coefficients the dashed vertical line marks
the null value. Comparing the posterior with the prior gives a visual
counterpart to the formal Bayes factors: posterior mass near zero
relative to the prior indicates little support for the effect, whereas
concentration away from zero indicates support for it.

\begin{figure}[htbp]
  \centering
  \includegraphics[width=\textwidth]{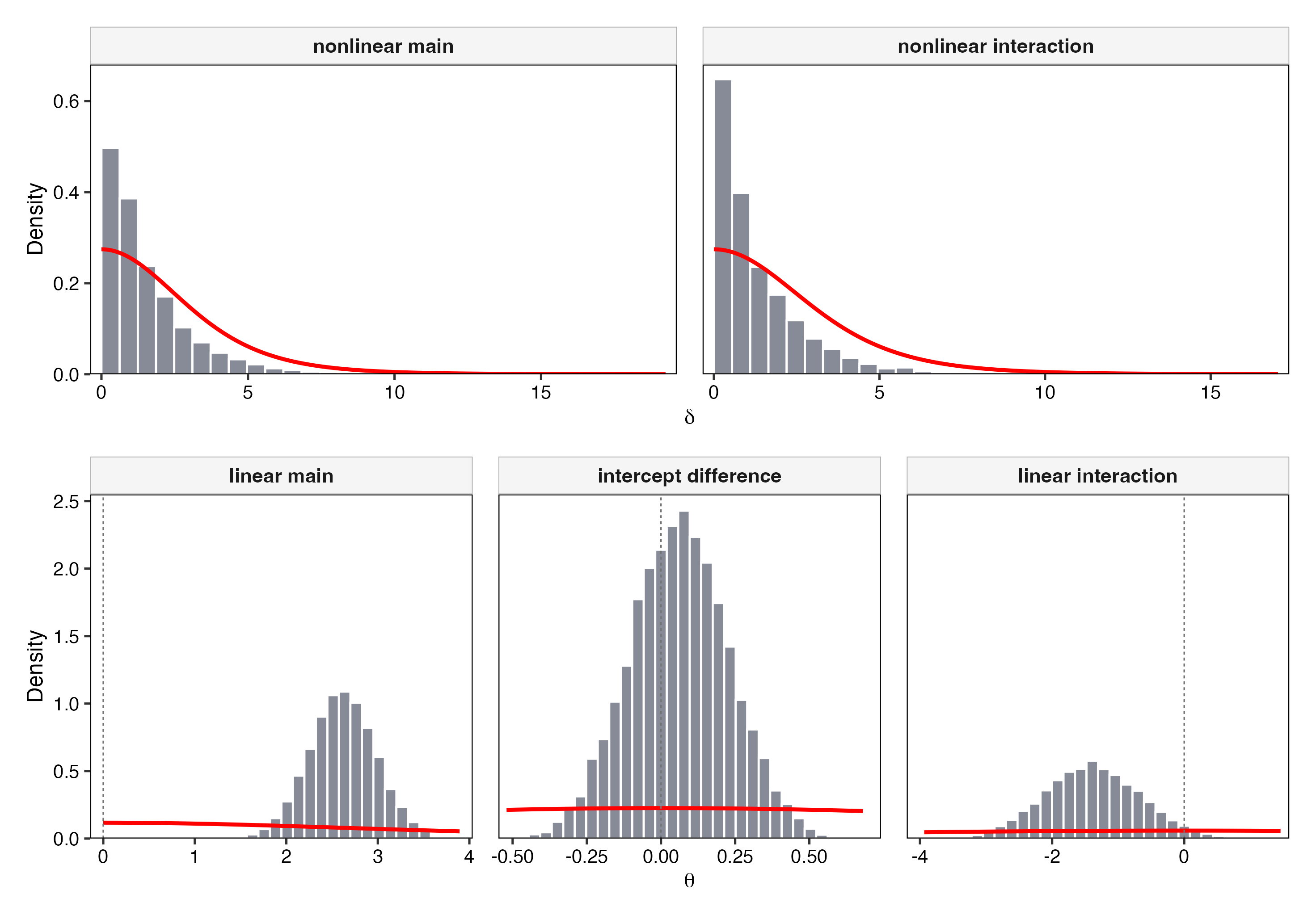}
  \caption{Mother's IQ application. Marginal posterior distributions
    (grey histograms) with the corresponding prior densities overlaid in
    red. The upper row shows the nonlinear magnitudes
    $\delta_{\text{main}}$ and $\delta_{\text{int}}$; the lower row
    shows the linear main effect $\beta_1$, the intercept difference
    $\beta_2$, and the linear interaction $\beta_3$, with the null value
    marked by a dashed vertical line.}
  \label{fig:post-a2}
\end{figure}

\begin{figure}[htbp]
  \centering
  \includegraphics[width=\textwidth]{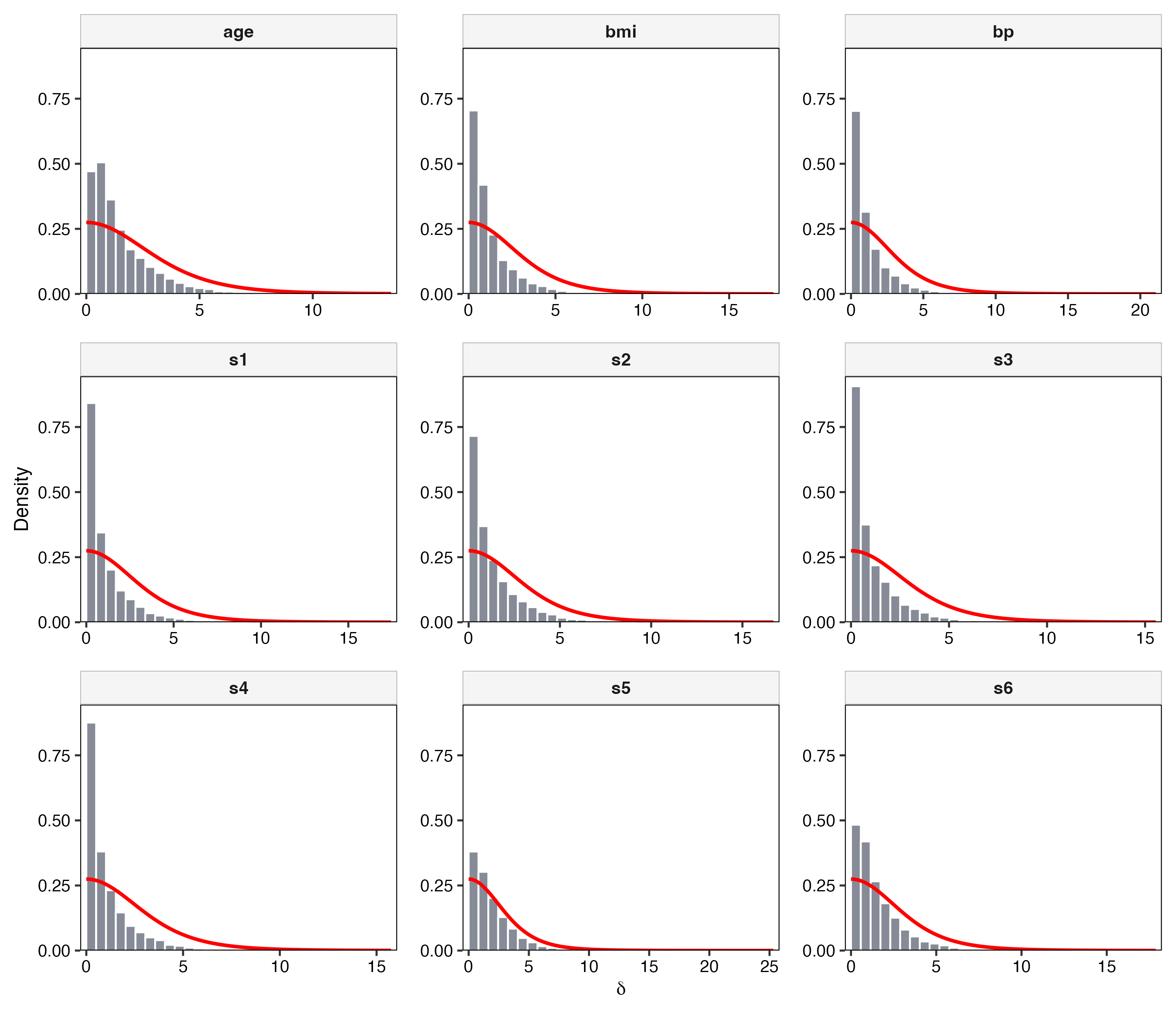}
  \caption{Diabetes dataset. Marginal posterior distributions of the
    nonlinear magnitude $\delta_j$ for each predictor (grey
    histograms), with the prior density overlaid in red.}
  \label{fig:post-a3-delta}
\end{figure}

\begin{figure}[htbp]
  \centering
  \includegraphics[width=\textwidth]{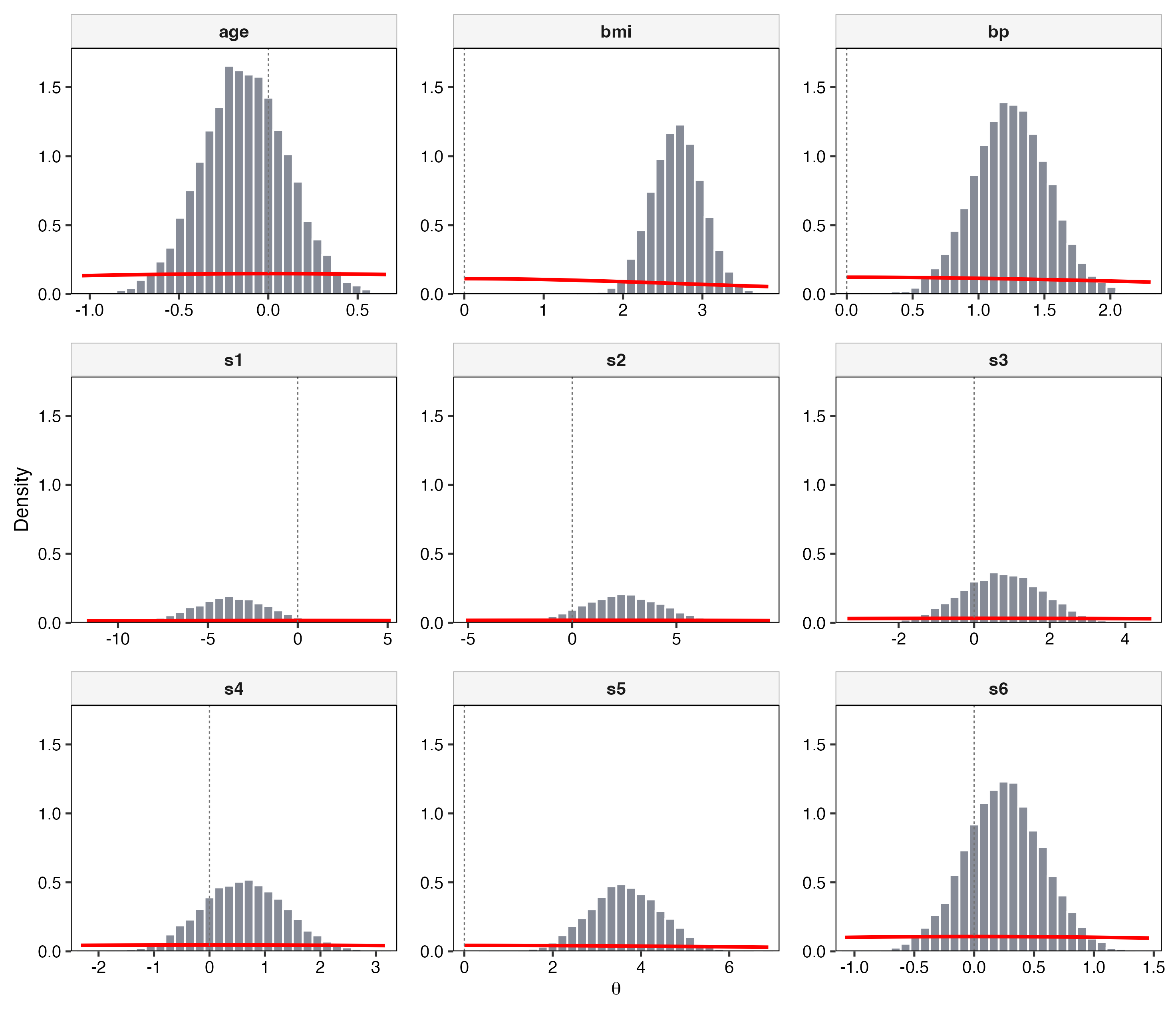}
  \caption{Diabetes dataset. Marginal posterior distributions of the
    linear coefficient $\theta_j$ for each predictor (grey
    histograms), with the prior density overlaid in red and the null
    value $\theta_j = 0$ marked by the dashed vertical line. }
  \label{fig:post-a3-theta}
\end{figure}

\begin{figure}[htbp]
  \centering
  \includegraphics[width=\textwidth]{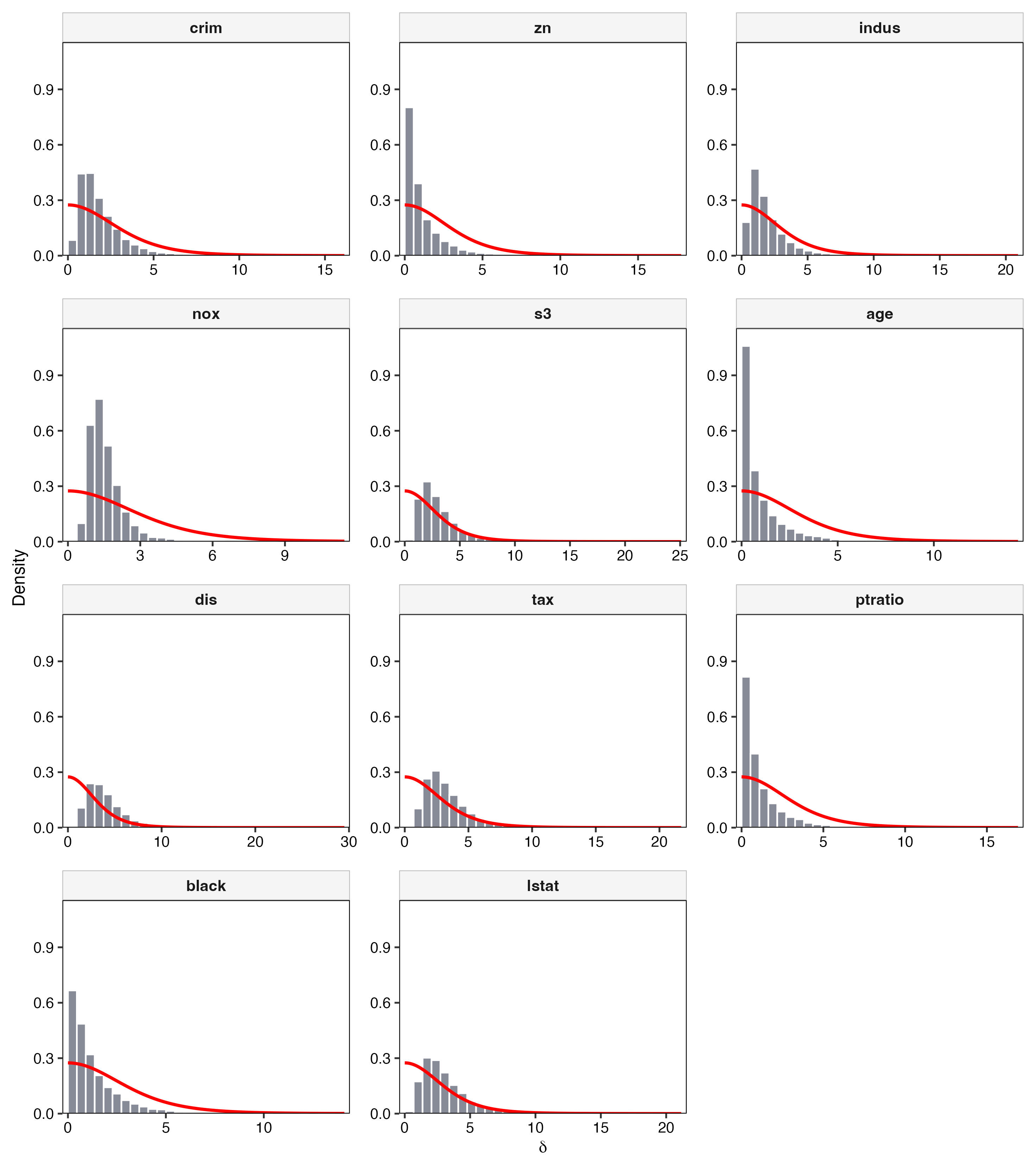}
  \caption{Boston housing dataset. Marginal posterior distributions of
    the nonlinear magnitude $\delta_j$ for each predictor (grey
    histograms), with the prior density overlaid in red.}
  \label{fig:post-a4-delta}
\end{figure}

\begin{figure}[htbp]
  \centering
  \includegraphics[width=\textwidth]{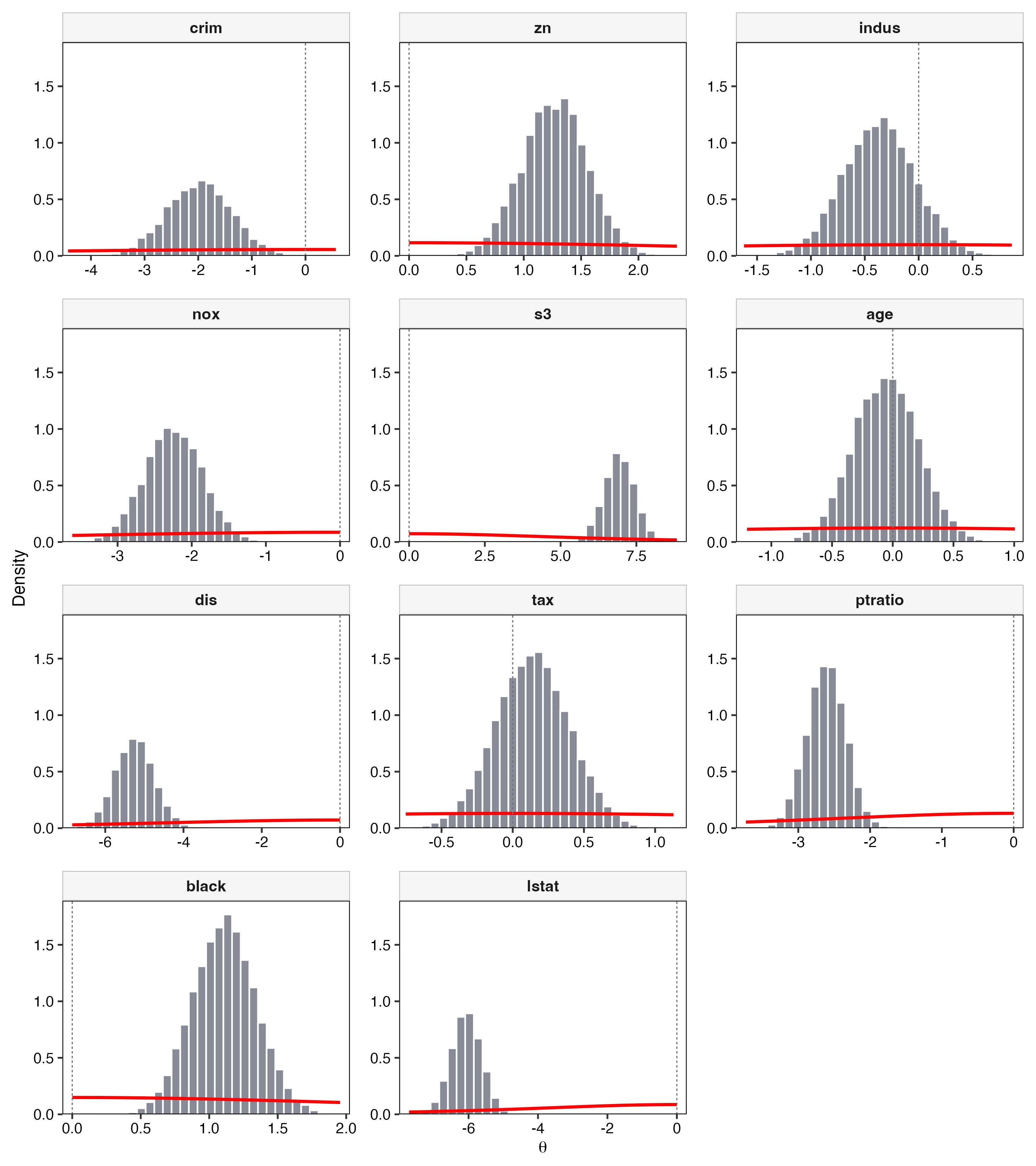}
  \caption{Boston housing dataset. Marginal posterior distributions of
    the linear coefficient $\theta_j$ for each predictor (grey
    histograms), with the prior density overlaid in red and the null
    value $\theta_j = 0$ marked by the dashed vertical line.}
  \label{fig:post-a4-theta}
\end{figure}

\newpage

\printbibliography

\end{document}